\documentclass[acmtog,authorversion=true,screen=true,authordraft=false,review=false,timestamp=false,anonymous=false]{acmart}

\AtBeginDocument{
  \providecommand\BibTeX{{
    \normalfont B\kern-0.5em{\scshape i\kern-0.25em b}\kern-0.8em\TeX}}}

\setcopyright{acmcopyright}
\acmJournal{TOG}
\acmYear{2020}
\acmVolume{39}
\acmNumber{4}
\acmArticle{125}
\acmMonth{7}
\acmDOI{10.1145/3386569.3392490}

\usepackage{caption}
\usepackage{subcaption}
\usepackage{enumitem}
\usepackage[nice]{nicefrac}
\usepackage{amssymb}
\usepackage{amsmath}
\usepackage{multirow}
\usepackage{dcolumn}
\newcolumntype{d}[1]{D{.}{.}{#1} }

\newcommand{\pare}[1]{{\left(#1\right)}}

\newcommand{\rem}[1]{}

\newcommand{\gc}{{g}}
\newcommand{\hc}{{h}}

\newcommand{\norm}[1]{\left\lVert#1\right\rVert}

\newcommand{\etal}{\textit{et al.~}}

\newcommand{\w}{\lambda}

\newcommand{\notchl}{l}

\newcommand{\matlab}{\textsc{Matlab}}
\newcommand{\tod}[1]{\overline{#1}}
\newcommand{\cutl}{{\mathcal{L}}}
\newcommand{\injR}{{ir}}
\newcommand{\Psur}{\mathcal{P}}
\newcommand{\Ppla}{\tod{\mathcal{P}}}
\newcommand{\cladf}{\mathcal{F}} 
\newcommand{\distf}{\mathcal{C}}

\begin{document}

\title[On Elastic Geodesic Grids and Their Planar to Spatial Deployment]{On Elastic Geodesic Grids and Their Planar to Spatial Deployment}

\author{Stefan Pillwein}
\email{stefan.pillwein@tuwien.ac.at}
\affiliation{
  \institution{TU Wien}
}
\author{Kurt Leimer}
\email{kurt.leimer@tuwien.ac.at}
\affiliation{
  \institution{TU Wien}
}
\author{Michael Birsak}
\email{michael.birsak@kaust.edu.sa}
\affiliation{
  \institution{KAUST}
}
\author{Przemyslaw Musialski}
\email{przem@njit.edu}
\orcid{0001-6429-8190}
\affiliation{
    \institution{NJIT and TU Wien}
}

\begin{abstract}

We propose a novel type of planar--to--spatial deployable structures that we call elastic geodesic grids. 
Our approach aims at the approximation of freeform surfaces with spatial grids of bent lamellas which can be deployed from a planar configuration using a simple kinematic mechanism. 
Such elastic structures are easy--to--fabricate and easy--to--deploy and approximate shapes which combine physics and aesthetics. 
We propose a solution based on networks of geodesic curves on target surfaces and we introduce a set of conditions and assumptions which can be closely met  in practice.  
Our formulation allows for a purely geometric approach which avoids the necessity of numerical shape optimization by building on top of theoretical insights from differential geometry.
We propose a solution for the design, computation, and physical simulation of elastic geodesic grids, and present several fabricated small-scale examples with varying complexity. 
Moreover, we provide an empirical proof of our method by comparing the results to laser-scans of the fabricated models. 
Our method is intended as a form-finding tool for elastic gridshells in architecture and other creative disciplines and should give the designer an easy-to-handle way for the exploration of such structures. 

\end{abstract}



\maketitle

\section{Introduction}

Design and construction of structures composed of curved elastic elements has a long history in the field of architecture. 
Alongside their aesthetical aspects imposed by nature, they have a lot of functional advantages: they are compact, lightweight and easy to build; nonetheless practicable, durable, and of high structural performance. 

They have been utilized for a long time dating back to ancient vernacular architecture for formal as well as for performance reasons, however, the possibilities of their form-finding in the past were limited \cite{Lienhard2013}. 
\vspace{40pt}

Fortunately, 
the currently available computational capabilities and advances in computer science open up avenues for direct modeling of complex shapes composed of elastically bending members. This goes beyond traditional architectural design and allows to aim at many general purpose products composed of such elements. The range of potential objects encompasses gridshells, formwork, paneling, various types of  furniture, sun and rain protectors, pavilions and similar small-scale buildings,  home decoration and accessories, like vases, bowls, or lamps, etc., and finally, also elements of future's functional digital fabrics that can be utilized in engineering as well as in fashion. 

\begin{figure}[t]
	\centering
	\includegraphics[width=1\columnwidth]{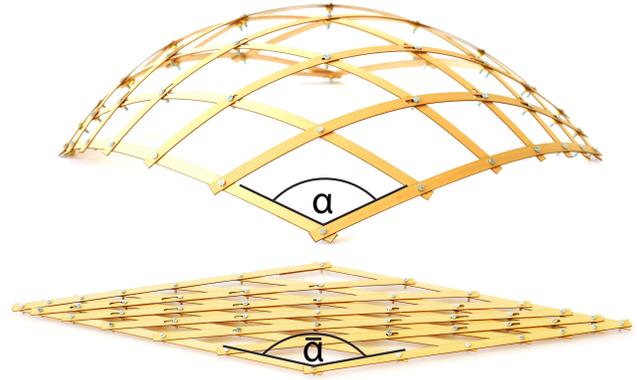}
	\caption{A deployed elastic geodesic gridshell (top) and its planar lattice in the rest state (bottom) fabricated of wooden lamellas. {The deployment of the whole kinematic system is based on changing angle $\tod{\alpha}$, such that $\tod{\alpha}\rightarrow\alpha$.  }}
	\label{fig:planar-deployed}
	\vspace{-1em}
\end{figure}

This vision leads directly to the objective of this paper: a designer provides a target surface and a computational method finds a planar grid of flat lamellas, that---when deployed---approximates the surface well. Figure \ref{fig:planar-deployed} shows a planar and a deployed grid of wooden strips, where a surface with the curved lamellas being tangential to it can be imagined.  
The joints between the lamellas allow for rotation and partially also for sliding. As the lamellas connecting opposite edges of the planar boundary quadrilateral are not parallel to each other, the grid is rigid in the plane. Given the flexibility of wooden lamellas with regard to bending and twisting, the grid is not rigid in space. By adjusting only one degree of freedom, for example the angle \begin{math} \tod{\alpha} \rightarrow \alpha \end{math} at one corner, the planar kinematic configuration \textit{elastically bends} continuously into a spatial gridshell which approximates the desired surface. The deployment process is governed by the rules of physics, seeing the lamellas as thin elastic minimal energy beams, allowed to bend as well as to rotate and slide at their intersections.

Our goal is to find a suitable planar setup of the lamellas that can be deformed into a spatial grid, fitting the target surface as closely as possible. 
To achieve this goal, we propose a solution based on networks of geodesic curves on the target surface. We introduce a set of conditions and assumptions which can be met closely in practice and restrict the grids to geodesics. 
However, at the same time, it allows us to develop a purely geometric solution which builds on top of theoretical background from differential geometry.

An advantage of our approach is to omit numerical shape optimization and to provide a solution which allows for easy exploration of designs of geodesic curve networks. 
To produce large scale gridshells, further considerations will be needed, however, our main goal is geometric modeling and form-finding. Our work provides insights into that domain, also due to the fact that it uses intrinsic surface geometry only. 
In summary, the contributions of this paper are the following: 
\begin{itemize}	
	
	\item We identify a specific case of the inverse design problem of spatial elastic grids which can be formulated using geometric considerations only. This formulation allows us to find a grid which is perfectly planar and can be isometrically deformed in an elasto-kinematic manner to a desired spatial grid. 
	
	\item We derive a mathematical method for form-finding of such geodesic grids based on differential-geometric properties of geodesic curves. In particular, we introduce \textit{distance maps} and \textit{cladding functions} which allow for efficient finding of suitable configurations without expensive numerical shape optimization. 
	
	\item Finally, we introduce physical simulation and a simple fabrication method for wooden small-scale elastic geodesic gridshells and perform empirical measurements which prove the validity of our approach. 
\end{itemize}

In the following section we review related work and in Section \ref{sec:definition} we provide a set of preliminary considerations necessary for our formulation. In Section \ref{sec:geometry} we provide the details of our geometric derivation, and in Section \ref{sec:model} we propose an adapted physical simulation. In Section \ref{sec:results} we present and evaluate our results. Finally, we discuss and conclude the work in Sections \ref{sec:discussion} and \ref{sec:end}.

\section{Related Work}\label{sec:related}

\paragraph{Developable Surfaces}
This topic has a long tradition in computer graphics and architectural geometry \cite{Pottmann2015a}. A lot of attention has been paid to the approximation of freeform surfaces with developable strips \cite{Wallner2010,Pottmann2010}, which can be fabricated from 2d flat material-sheets by cutting. By bending and combining them, complex freeform surfaces can be erected. 
Also paneling of surfaces with planar tiles \cite{Eigensatz2010b} or with general planar polygons \cite{Chen2013} have been proposed. Another way is the division of shapes into principal strips which bend automatically if combined \cite{Takezawa2016a}. 
On the theoretical side, a novel representation of developable surfaces using quadrilateral meshes with appropriate angle constraints~\cite{Rabinovich2018} or a definition of developability for triangle meshes~\cite{Stein2018} have been proposed recently. Also discrete geodesic parallel coordinates for modeling of developable surfaces were proposed~\cite{Wang2019}. 
All these works aim at the design of developable surfaces, which, due to their isometric properties, can be fabricated from planar sheets. However, they do not incorporate a planar-to-spatial elastic deployment.

\paragraph{Deployable Surfaces}
One more way to easily construct spatial shapes from flat sheets is by appropriately folding paper \cite{Mitani2004b,Massarwi2007a}, which is inherently related to the Japanese art of Origami \cite{Dudte2016}. 
Another set of works deals with curved folding and their efficient actuation from flat sheets to spatial objects \cite{Kilian2008,Kilian2017}. 
Our work is related to these approaches in terms of being deployable from a planar initial state, however, the main difference is that our grids are elastic and approximate doubly-curved surfaces.  

In fact, a lot of attention has been paid to the design of doubly-curved surfaces which can be deployed from planar configurations due to the ease of fabrication. One way of achieving this goal is by using auxetic materials \cite{Konakovic2016} which can nestle to doubly-curved spatial objects, or in combination with appropriate actuation techniques, can be used to construct complex spatial objects \cite{Konakovic-Lukovic2018}. The main difference to our approach is that these structures do not use elastic bending to reach the actual spatial shape.

\paragraph{Elastically Deployable Surfaces}
An interesting way to deploy surfaces is to utilize the energy stored in planar configurations in order to approximate shapes, for instance using prestressed latex membranes in order to actuate precomputed planar geometric structures into freeform shapes \cite{Guseinov2017}, or to predefine flexible micro-structures which deform to desired shapes if set under tension \cite{Malomo2018a}. A combination of flexible rods and prestressed membranes lead to Kirchhoff-Plateau surfaces that allow easy planar fabrication and deployment \cite{Perez2017a}. 
These methods achieve their planar-to-spatial configuration from elastic tension in the network, either due to prestressing in the planar state or by setting appropriate boundary conditions. The latter approach is more closely related to ours, however, instead of structure optimization, we build on top of the differential geometric properties of geodesic curves on the target surfaces. Thus, our method is based on the assumption that the elastic elements can bend and twist, but not stretch and must therefore maintain the same length in the planar as well as in the spatial configuration.

\paragraph{Wire Surfaces}
Our work also contributes to surface approximations using grids. This is not a novel approach, and previous works have tackled this topic. 
For example, approximations of surfaces with meshes based on Chebyshev nets \cite{Garg2014}, as well as with wires that are deformed in planar configurations and assembled together \cite{Miguel2016} to abstract a spatial shape, have been proposed.  
In contrast to us, these works do not focus on elastic-planar-to-spatial deployment nor on elasticity of the networks.

\paragraph{Physical Surfaces. }
A number of methods which aim directly at computational design of physically valid and stable architectural structures have been proposed. For example, design of self-supporting masonry surfaces \cite{Vouga2012a} or the design of unreinforced masonry surfaces \cite{Panozzo2013a}. Also the process of erection of such objects has been computationally explored \cite{Deuss2014}. Moreover, methods for fast interactive form-finding of physically stable structures \cite{Tang2014a}, for the minimization of material usage under stability constraints \cite{Kilian2017a}, or physically plausible tensegrity structure design~\cite{Pietroni2017} have been proposed. 
Our method is related in terms of the goal of achieving structurally stable shapes. In turn, these methods do not utilize elastic bending for deployment or stability.

\paragraph{Classical Geometric Surfaces.}
In classic differential geometry,  geodesic nets on surfaces which can be mapped onto a geodesic net on a different surface (including a plane) have been analyzed by Voss \shortcite{Voss} and Lagally \shortcite{Lagally}. Regarding to their analysis, arc-length preserving mappings of continuous geodesic nets onto each other require rhombic geodesic nets, i.e., need a parametrization of the surface with the net curves as parameter curves and \begin{math} E = G \end{math} in the fundamental form. The resulting  Liouville surfaces are very limited in shapes, and therefore not useful for our freeform design purpose.

\paragraph{Gridshells and Active-Bending}
The idea of grid\-shells---{struc\-tures that gain their strength and stiffness through their curvature
---were introduced by Shukhov for the Rotunda of the Panrussian Exposition \cite{Shukhov1896} and further pursued by famous architects, e.g., by Frei Otto for the construction of the roof of the Multihalle at the Mannheim Bundesgartenschau \cite{Happold1975}. 

The introduction of the \textit{active bending} paradigm \cite{Lienhard2013} together with enhanced and easy-to-use computational methods increased the interest of the scientific community in systematically utilizing elastic bending to realize curved shapes. Until recent advances in computer science they could only be form-found empirically \cite{gengnagel2013active}. 

Existing design approaches are often based on particular kinds of surface curves, e.g., curvature lines \cite{Schling2018}. Emerging concepts for the  erection of elastic gridshells facilitate the construction process or even eliminate the need for scaffolding \cite{Quinn2014}. 

Architectural works which aim at the approximation of gridshells and combine lightweight structural design with aesthetics \cite{Soriano2017} also inspired our work. 
Soriano et al. \shortcite{Soriano2019} also proposed mechanisms for the deployment of geodesic gridshells using an evolutionary solver to form-find the grids. 
However, the design process is rather complex and time consuming, using numerical gradient-free optimization methods. In contrast, our approach is based on geometric considerations and omits expensive computations. 
Besides gridshells, kinetic structures, bending plate structures, and textile hybrids form a new class of structures explored in the active bending research community \cite{Lienhard2018}.

Recently \cite{Panetta2019} introduced an interactive approach for finding deployable grid structures. Their method requires the user to create an initial grid design by iterating between layout editing and grid simulation steps. Once an overall satisfying shape is found, the layout is then optimized to reduce the internal elastic energy of the flat assembly state and the deployed target state.

In contrast, our design approach only requires the user to provide a target surface patch. Based on its geometry, our algorithm produces a grid layout to approximate the target surface patch when deployed. 
Furthermore, our approach guarantees that the planar configuration is in a zero-energy state.

\begin{figure}[t]
	\centering
	\includegraphics[width=1.0\columnwidth]{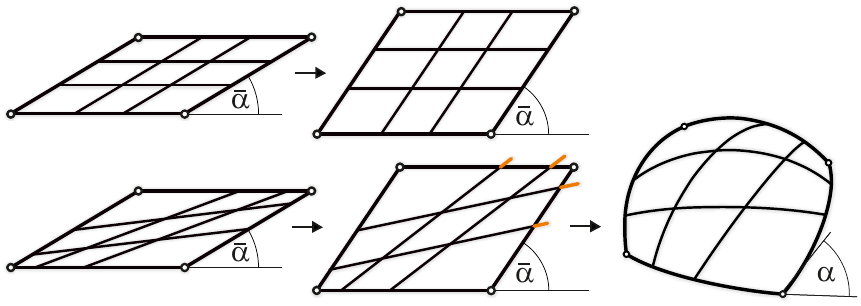}
	\caption{The principle behind our planar to spatial deployment system. Top row: all members of a family are parallel and rigid, the kinematic linkage can move freely in the plane. Bottom row: non parallel layout produces a deadlock when trying to change the shape, inner members are too long. Allowing members to elastically deform, they buckle out of plane.}
	\label{fig:kinematics}
	\vspace{-5pt}
\end{figure}

\paragraph{Fabrication and Elastic Simulation}
The computer graphics community started to deal with fabrication and computational design \cite{Bermano2017}, for this reason many novel methods aim at fast but physically valid simulations. Our simulation is based on the method of discrete elastic rods \cite{Bergou2008,Bergou2010}, which have been adapted and utilized for works on sparse rod networks \cite{Perez2015,Malomo2018a,Vekhter2019}. Recently this method has been also used for the simulation of hemispherical elastic gridshells \cite{Baek2019}.

\section{Preliminary Considerations}\label{sec:definition}
\subsection{Elasto-Kinematic Deployment}\label{sec:problem}

The main idea behind our pla\-nar-to-spatial deployment is based on a very simple kinematic mechanism, as depicted in Figure \ref{fig:kinematics}. It is a special case of a planar quadrilateral four-bar linkage with rigid members, rotating joints and one degree of freedom. 

If we change the angle at one corner and all links of a family are parallel, the system can move freely in the plane (Figure~\ref{fig:kinematics}, top row).
If we introduce stiff inner links which are not parallel, the system is deadlocked. By introducing bending and twisting flexibility to the members, they buckle out of plane in order to preserve their length and form a spatial grid (Figure~\ref{fig:kinematics}, bottom row).
To construct such a mechanism, the lengths of the members must match on the surface as well as in the planar configuration. 
Mathematically, this behavior can be modeled by geodesic curves on a surface.

A geodesic locally minimizes the arc length between two distinct points and maintains its length under isometric deformations of the surface. 
Moreover, its principal normal falls into the surface normal, i.e., it allows normal curvature, but prohibits geodesic curvature. 
As a consequence, a carefully chosen network of such curves can be used to build the elasto-kinematic deployment mechanism and at the same time to abstract the surface' characteristics. 

Additionally, gridshells of the nets should be easy to manufacture, transport, assemble, and deploy. 
To meet these properties in practice, we use thin straight lamellas with a cross section ratio of about $1 : 10$, creating a distinct weak axis for easy bending and a strong axis that prohibits bending. These lamellas can be wrapped on a surface and interpreted as tangential strips with a geodesic centerline. Also their connections, which are essential for the kinematic deployment, imitate the intersections of geodesics well: the lamellas can rotate with the axis of rotation being always parallel to both of the principal normals of the centerlines, and their connections can slide along the tangents of the centerlines. 

Besides apparent advantages of easy production, geodesics offer a lot of theory and give us a great set of tools to analyze surface patches and find suitable solutions.

\begin{figure}[t]
	\centering
	\includegraphics[width=1.0\columnwidth]{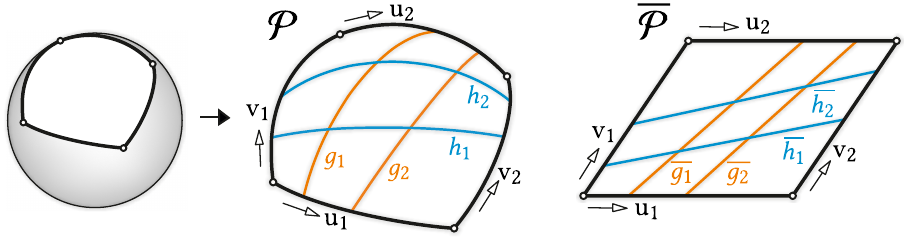}
	\caption{Overview of our approach and the notation. Left: the user selects four corners on a desired target surface. Center: the surface patch $\Psur$ with members of the $\gc$ and $\hc$ family. Each family is parameterized with pairs $(u_1,u_2)$ and $(v_1,v_2)$ respectively. Right: a corresponding planar patch $\Ppla$ with corresponding members of the $\tod{\gc}$ and $\tod{\hc}$ family (cf. Section~\ref{sec:representation}). }
	\label{fig:overvew2}
	
\end{figure}

\subsection{Grid Representation}\label{sec:representation}
The input to our computational system is a surface patch $\Psur$ which is a convex bounding shape defined on a designer created target surface by four corners. 
They are connected by geodesic curves on the surface which constitute the boundaries of the surface patch $\Psur$ as depicted in Figure~\ref{fig:overvew2}. 
The output of our system is a planar quadrilateral, denoted as planar patch $\Ppla$, filled with interconnected straight lines. 
Its corners are the counterparts of the spatial corners.

The patches consist of two families of grid members: ${\gc, \hc}$-members are geodesics on the surface patch, and ${\tod{\gc},\tod{\hc}}$-members are their corresponding straight lines in the planar patch with matching lengths (cf. Figure \ref{fig:overvew2}). 
The grid members are parameterized along the boundaries with parameter-pairs $\pare{u_1,u_2}$ and $\pare{v_1,v_2}$ respectively.

\subsection{Surface Patch Characteristics}\label{sec:patch}

Using geodesics to model the grid members also poses restrictions on the representability of the target surfaces. 
There are two ways to compute geodesics: defining a start point and a direction vector, which has a unique solution, or defining a start and an end point, which delivers the shortest path between these two points, but does not necessarily have a unique solution \cite{Polthier1998}.

To maintain the length of a curve between the boundaries, we need to compute geodesics between two points on opposite boundaries, so for our application we use the second case, which we will denote as shortest geodesics from now on.

A feature of shortest geode\-sics---namely the possibility of non unique solu\-tions---can have disadvantageous effects for the approximation. It may happen that two points on a surface patch can be connected by more than one shortest geodesic. The existence of such points is linked to the Gaussian curvature $K$ of the surface. They result in areas of the patch $\Psur$ that cannot be covered with shortest geodesics connecting the boundaries. For the quality of the approximation, it needs to be ensured that every point on patch $\Psur$ can be reached by a shortest geodesic of the $\gc$ and $\hc$-curves family. If this is not the case, surface features cannot be captured with shortest geodesics and cannot be encoded in the planar grid.

Figure \ref{fig:cladding} illustrates the problem: when drawing shortest geodesics from point $p$ to all points on the opposite boundary, the central area of high positive $K$ remains uncovered and produces a gap in the coverage. 
Taking a look at the distance field (Figure~\ref{fig:cladding}, left), we can identify singularities  as it approaches the opposite boundary. These singularities form the \emph{cut locus} $\cutl(p)$ on $\Psur$ and each point $\in \cutl(p)$ can be reached from $p$ by two distinct geodesics of the same length. 

\begin{figure}
	\centering
	\includegraphics[width=1\columnwidth]{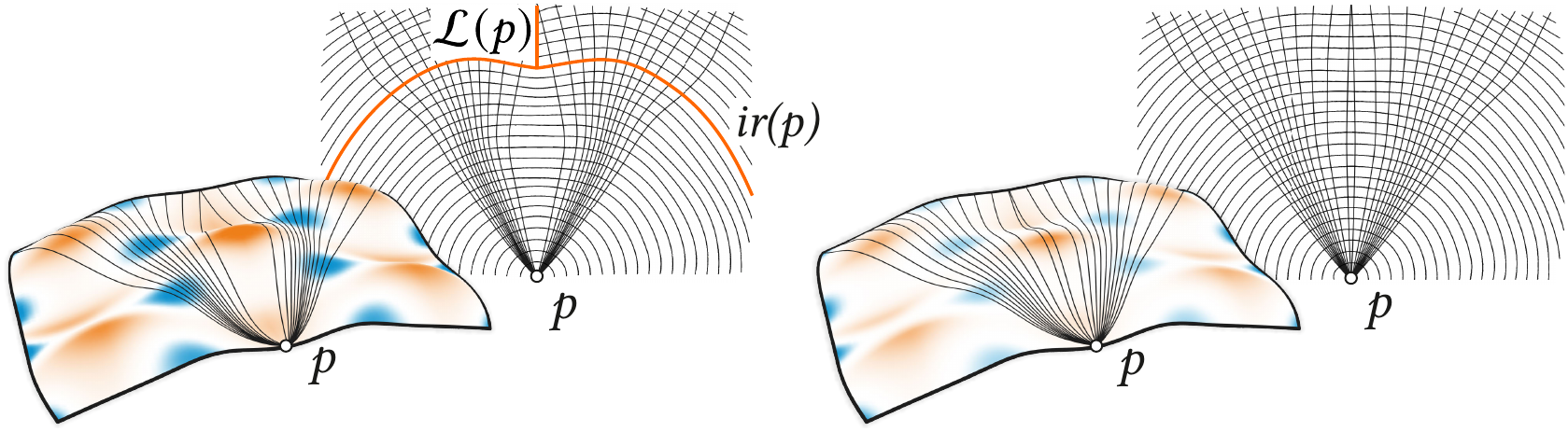}
	\caption{Shortest geodesics between point $p$ and points on the opposite boundary (top) and distance fields emanating from $p$  (bottom). Left:  the peak area cannot be covered by shortest geodesics, cut locus $\cutl(p)$ and injectivity radius $\injR(p)$ are indicated. Right: Uncovered area sufficiently reduced by smoothing (cf. Section~\ref{sec:patch}).} 
	\label{fig:cladding}
\end{figure}

The geodesic distance $d$ between $p$ and its nearest point on $\cutl(p)$ is called the \emph{injectivity radius} $\injR(p)$  \cite{Carmo1992} given as 
\begin{equation*}
\injR(p) = \inf~d(p,\cutl(p)) \,. 
\end{equation*} 

Using a corollary of the Rauch comparison theorem \cite{Carmo1992} we obtain the following inequality: 
\begin{equation} 
\label{eq:injectivity} 
\injR(p) \ge \frac{\pi}{\sqrt{K_\text{max}}} \,.
\end{equation}
It gives us a lower bound for the injectivity radius $\injR(p)$ for each surface point $p$. 
Evaluating it at local peaks of Gaussian curvature $K_\text{max}$ serves as a quick check for the uniqueness of shortest geodesics.

If the lengths of all members are smaller than the right hand side of Expression (\ref{eq:injectivity}), the patch can be used as it is. 
If this is not the case, the surface patch cannot be covered completely (unless the peak is on the boundary).

Although Expression (\ref{eq:injectivity}) indicates the existence of these areas, the size of the gaps remains unclear. 
Small gaps may not pose big problems for the quality of the approximation, while big gaps do.
They indicate that there is a considerable difference in length between the shortest geodesic next to the peak and the (start-direction) geodesic over the peak, thus 
the quality of the approximation of the surface by the planar grid will be worse. 
In order to handle surface patches that cannot be covered with shortest geodesics completely, we propose an iterative smoothing procedure. 

To check for uncoverable areas around a Gaussian curvature peak $p_\text{max}$, we first compute two distance fields: one from the peak $p_\text{max}$ and one from the boundary point $p_1$, where we choose $p_1$ to be the closest point to  $p_\text{max}$ on the boundary. 

They provide us with distances $d(p_1,q)$ to the points $q$ of the opposite boundary as well as $d(p_1,p_\text{max})$ and $d(p_\text{max},q)$. We compute the minimum of $d(p_1,p_\text{max}) + d(p_\text{max},q) - d(p_1,q)$, which is reached at a point $q_1$. If the minimum is close to zero, the peak $p_\text{max}$ is not problematic and there is no gap. If not, the factor:
\begin{displaymath}
\eta = \frac{d(p_1,p_\text{max}) + d(p_\text{max},q_1)}{d(p_1,q_1)}
\end{displaymath}
is used to measure the size of the gap. 
In order to remove the unreachable gaps, we perform Laplacian smoothing of $\Psur$ with cotangent weights iteratively \cite{Desbrun1999a}, until $\eta$ falls below a certain threshold $\eta_\text{max}$. 
In practice we choose $\eta_\text{max} = 1.0015$ (cf. Figure \ref{fig:cladding}, right) which we have determined empirically.

\section{Elastic Geodesic Grids}\label{sec:geometry}

\begin{figure}[t]
	\centering
	\includegraphics[height=32.2mm]{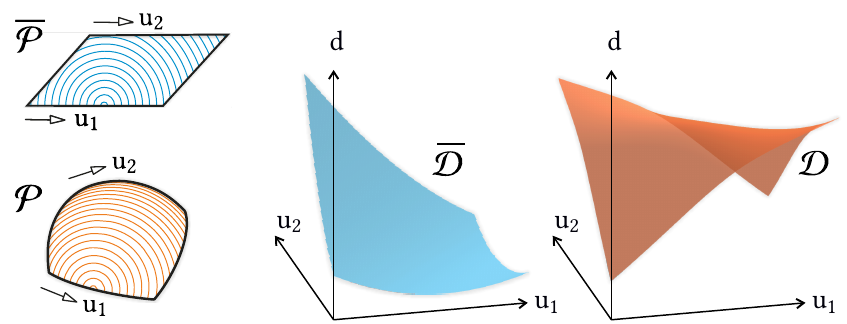}
	\caption{Distance fields on a planar patch $\Ppla$ and a surface patch $\Psur$, computed from a single point shown on the left. By sampling all point-pairs along corresponding $(u_1,u_2)$-domains, we create distance maps $\mathcal{D}_u(u_1,u_2)$ and $\tod{\mathcal{D}}_u(u_1,u_2,\tod{\alpha})$. Note that the planar distance map $\tod{\mathcal{D}}$ also depends on the shape of $\Ppla$ and thus the angle $\tod{\alpha}$ (cf. Section~\ref{sec:distmaps}).  } 
	\label{fig:dFields}
	\vspace{-5pt}
\end{figure}

\subsection{Grid Criteria}\label{sec:criteria}

Our goal is to find a grid of geodesics on $\Psur$, which can be ``planarized'' to $\Ppla$ with a certain angle $\tod{\alpha}$.
The grid curves are allowed to reduce their curvature and torsion but should keep their total lengths as well as the lengths between points of intersection.
At each configuration, the grid curves should be geodesics on a hypothetical surface. 

Inversely, the planar grid is deployed to a spatial grid as the planar angle approaches the spatial angle, i.e.,  $\tod{\alpha}\rightarrow\alpha$ such that the planar corners approach their spatial counterparts, and the planar straight lines bend to geodesic curves tangential to the target surface. 

In order the meet these requirements, both the planar and the spatial grids need to obey the following geometric demands:
\begin{enumerate}[label=(\roman*)]
	\item\label{GC1}  \emph{Length correspondence}: All straight lines $ \tod{\gc}, \tod{\hc} $ have the same lengths as their corresponding geodesics  $ \gc, \hc.$
	
	\item\label{GC2}  \emph{Boundary correspondence}:  On boundaries, the $(u_1,u_2)$ and $(v_1,v_2)$ coordinates of connections are identical for the 2d and the 3d grid.  
	
	\item\label{GC3} \emph{Bijectivity of correspondence}: Each point on one boundary has one and only one corresponding point on the opposite boundary, defining a grid member uniquely.
	\item\label{GC4} \emph{Convexity of boundary}: the corresponding patches $\mathcal{P}$ and $\tod{\mathcal{P}}$ need to be convex. 
	
\end{enumerate}
Criterion~\ref{GC4} is necessary, since otherwise the kinematic mechanism can run into a deadlock. 
It is fulfilled if each of the four inner angles of $\Psur$ is less than $\pi$, which can be argued with the triangle inequality of the surface metric and the convexity of sufficiently small areas \cite{Carmo1992}.

In the following, we introduce mathematical tools which allow to identify geodesic grids which fulfill all posed criteria. 
We explain the process only for one family of members.
Note however that the shape of the planar patch is chosen with respect to both families, satisfying interconnecting constraints, thus they are not found independently.

\subsection{Distance Maps}\label{sec:distmaps}

\begin{figure}[t]
	\centering
	\includegraphics[height=32.2mm]{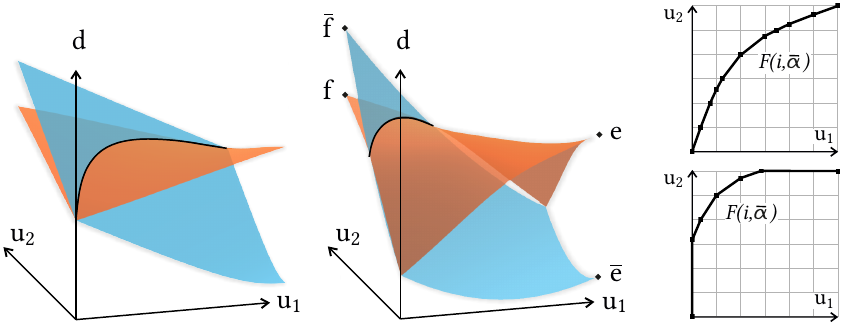}
	\caption{Intersection of distance map $\tod{\mathcal{D}}_u(u_1,u_2, \tod{\alpha})$ for planar patch in blue and distance map $\mathcal{D}_u(u_1,u_2)$ for surface patch in orange. Left: proper intersection, fulfilling the constraints (cf. Sec.~\ref{sec:cladding}). Center: partial intersection, providing an invalid cladding function $\cladf_u$. Right: piecewise linear functions $F_u$ of both cases evaluated on a discrete grid (cf. Section~\ref{sec:cladding}). 
	}
	\label{fig:distanceMaps}
	\vspace{-1em}
\end{figure}

As a tool to match the distances on the surface patch $\Psur$ and the planar patch $\Ppla$, we introduce distance maps $\mathcal{D}_u$ and $\mathcal{D}_v$. To create them, distance fields are spread from all points $p(u_1)$ on one boundary to all points $q(u_2)$ on the opposite boundary, measuring the geodesic  distances $d(p(u_1),q(u_2))$ between them (cf Figure \ref{fig:dFields}, left). Transforming the distances into the $\pare{u_1,u_2, d}$-3d space creates a representation of the geodesic lengths of the surface patch, which is illustrated in Figure \ref{fig:dFields}. While the distance maps of the surface patch $\mathcal{D}_u(u_1,u_2)$ and $\mathcal{D}_v(v_1,v_2)$ have a predefined angle $\alpha$ induced by the choice of the surface patch and depend only on the coordinates $u_1,u_2$ and $v_1,v_2$ respectively, the distance maps of the planar patch $\tod{\mathcal{D}}_u(u_1,u_2, \tod{\alpha})$ and $\tod{\mathcal{D}}_v(v_1,v_2,\tod{\alpha})$ also depend on the angle $\tod{\alpha}$. The choice of that  angle changes the shape of the planar grid and hence also the shapes of the distance maps $\tod{\mathcal{D}}_u$ and $\tod{\mathcal{D}}_v$.

In our implementation, distance maps are represented as quad meshes; their resolution is chosen according to the resolution of the input surface mesh. In practice, it is around $100\times100$ vertices.

\subsection{Cladding Functions}\label{sec:cladding}

In this section we derive the cladding functions which determine the distribution of the corresponding members in $\Ppla$ and $\Psur$. 
This is done via finding a suitable angle $\tod{\alpha}$, such that the grid criteria defined in Section \ref{sec:criteria} are fulfilled.  

\begin{figure*}[t]
	\centering
	\includegraphics[width=1\textwidth]{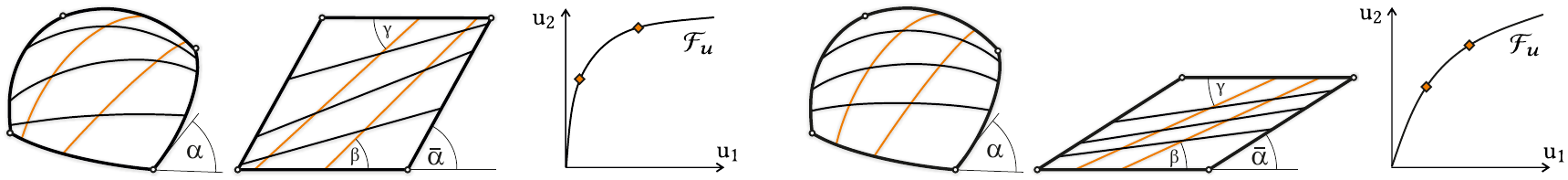}
	\caption{
		The influence of $\tod{\alpha}$ on the cladding with grid members: its choice affects the distribution and coverage of the members $\gc$ and $\hc$ on the surface patch $\Psur$. Right: the shape of the cladding function $\cladf_u$ with indicated members (cf.~Section~\ref{sec:cladding}). Please note also  the angles $\beta$ and $\gamma$, which are used to determine minimum distances between lamellas with a certain width (cf.~Section~\ref{sec:members}). }
	\label{fig:alpha}
	\vspace{-6pt}
\end{figure*}

The cladding function $\cladf_u$ is built by first projecting the intersection of the distance maps $\mathcal{D}_u$ and $\tod{\mathcal{D}}_u$ to the $u_1, u_2$-plane (respectively,  $\cladf_v$ is built using a projection to the $v_1, v_2$-plane). 
Points on this function represent geodesics which connect opposite boundaries and have the same length on both the planar and the spatial patch.  
Please recall that the shape of the distance map $\tod{\mathcal{D}}_u(u_1,u_2, \tod{\alpha})$ also depends on the choice of the angle $\tod{\alpha}$, hence the shape of the cladding function does as well.

Grid criteria~\ref{GC1} and~\ref{GC2} are fulfilled by the nature of these functions. 
Our goal is now to determine the parameter $\tod{\alpha}$ such that also grid criteria~\ref{GC3} and~\ref{GC4} are fulfilled. This implies that the cladding function ${\cladf}_u$ must be continuous and bijective over the entire domain, which means its first order partial derivative $\dot{\cladf}_u$ w.r.t. $u_1$ should nowhere reach $0$ nor $\infty$ (cf. Figure~\ref{fig:distanceMaps}, right). 

Additionally, bounds can be set on $\dot{\cladf}_u$ in order to avoid too steep or too flat tangents, which would result in a strong concentration of members on a boundary and an uneven coverage of the patches  $\Ppla$ and $\Psur$ as shown in Figure \ref{fig:alpha}. 
Moreover, if criteria~\ref{GC3} and~\ref{GC4} are not fulfilled, triangular member connections may appear in the planar grid, destroying the kinematic deployment mechanism.

With this picture in mind, we denote the cladding functions as
\begin{equation*} 
u_2 = \cladf_u(u_1,\tod{\alpha})  \quad\text{and}\quad v_2 = \cladf_v(v_1,\tod{\alpha})  \,
\end{equation*}
with $u_1, u_2 \in [0, 1]$  ($v_1, v_2$ respectively). 
Refer to Figure~\ref{fig:alpha} for a depiction. 
Please note that for the cladding functions to exist, the length of the diagonals $e,f$ of the surface patch $\Psur$ and $\tod{e},\tod{f}$ (cf. Figure~\ref{fig:distanceMaps}) of the planar patch $\Ppla$ must fulfill the following inequality: 
\begin{equation} \label{eq:diagonals} 
(e-\tod{e}) \cdot (f-\tod{f}) < 0  \,.
\end{equation}
In other words, this inequality is a necessary condition for a proper intersection of the distance maps. 
Figure \ref{fig:distanceMaps} depicts how the diagonals $e,f$ of the surface patch and $\tod{e},\tod{f}$ of the planar patch appear in the distance maps.

To find a feasible domain for the angle $\tod{\alpha}$ under the condition of bijective cladding functions $\cladf_u(u_1,\tod{\alpha})$ and $\cladf_v(v_1,\tod{\alpha})$, we formulate it as an optimization problem using Expression (\ref{eq:diagonals}) as a constraint. 

Note that at $(0,0)$ and $(1,1)$ distance maps always intersect, so $\cladf_u$ is always defined there. However, the function might be not defined or not continuous over the entire domain of $u_1 \in [0,1]$, as depicted in Figure~\ref{fig:distanceMaps}, center. 
To deal with this case, we introduce a piecewise linear parametric representation ${F}_u(i, \tod{\alpha}) = (u_1(i), u_2(i), \tod{\alpha})$ given over the entire domain and range of $\cladf_u$ (cf. Figure~\ref{fig:distanceMaps}, right).

Using the slopes of the segments $\dot{F}_u$ and $\dot{F}_v$ simultaneously as constraints, we cast the following optimization problem to determine a feasible domain for the angle: 
\begin{equation}\label{eq:alpha}
\begin{aligned}[l,l]
&\min &&~ \tod{\alpha} \\
&\text{s.t.} && (e-\tod{e}) \cdot (f-\tod{f}) < 0 \\
& &&  k_\text{min} < \dot{F}_u(i, \tod{\alpha}) < k_\text{max}, \;\; 1\dots n   \\
& &&  k_\text{min} < \dot{F}_v(i, \tod{\alpha}) < k_\text{max}, \;\; 1\dots n, 
\end{aligned}
\end{equation}
with $n$ being the number of segments and with $k_\text{min}$ and $k_\text{max}$ being slope bounds which we have determined empirically as $k_\text{min} = 0.1$ and $k_\text{max} = 10$. 
We evaluate $\dot{F}_u$, $\dot{F}_v$ using finite differencing 
\begin{equation*} 
\dot{F}_u(i, \tod{\alpha}) = \frac{\Delta u_2(i)}{\Delta u_1(i)}
\end{equation*}
at all segments, as shown in Figure~\ref{fig:distanceMaps}, right. 
To tackle the case where $\dot{F}_u = \infty$, we set its value to $c\Delta u_2$ with $c\gg k_{\text{max}}$; cases with $\dot{F}_u = 0$ do not cause any numerical problems. 
In our implementation, each cladding function is computed by intersecting the distance map meshes and their resolution induces the resolution of piecewise linear function $F$.

We solve Problem (\ref{eq:alpha}) using sequential quadratic programming with numerical gradients w.r.t. $\tod{\alpha}$.
First we determine the minimum feasible $\tod{\alpha}_{\min}$ with the lower bound for $\tod{\alpha}$ from the convexity restrictions of grid criterion~\ref{GC4}. 
Then we find a maximum feasible $\tod{\alpha}_{\max}$ using the same concept.  
Values of $\tod{\alpha}$ between these bounds ensure the cladding functions $\cladf_u$ and $\cladf_v$ to be bijective. 

Note, that setting bounds for $\tod{\alpha}$ also makes it possible to introduce designer constraints on the shape of the planar patch $\Ppla$. In practice, we choose $\tod{\alpha}_{\min}$ for our examples, which results in a compact planar patch design.

\subsection{Grid Members}\label{sec:members}

After checking the validity of the surface patch (with smoothing, if needed) and fixing $\tod{\alpha}$,  we choose the number and positions of the grid members. 
Patches with many curvature features (compare Figure \ref{fig:cladding}) obviously need a minimum number of well placed members to capture all surface features well. For this specific example, all the bumps of the surface have to be encoded in the planar grid.

Our approach for fitting grid members  is a geometrically motivated heuristic. It reuses the information from the intersections of the respective distance maps  $\mathcal{D}_u$ and $\tod{\mathcal{D}}_u$ in the $(u_1$, $u_2$, $d)$ space (cf. Section \ref{sec:cladding}). 
Along their intersection curve, we can construct an associated function $\distf_u(s)$ of geodesic lengths $d$ of the members. 
Its maxima and minima correspond to longest or shortest geodesics ($g_i, \tod{g_i}$) on the surface patch $\Psur$ and provide good candidates for physical members of the elastic grid. 

Hence, members are first placed at the extrema of $\distf_u(s)$ and next at the extrema of the curvature of $\distf_u(s)$. 
The first pass ensures to cover major features  (large peaks) since these members correspond to locally longest and shortest geodesics. The second pass ensures to capture finer features (smaller bumps), since the corresponding members are also locally the longest or the shortest members, however on a smaller scale. Figure~\ref{fig:densify} depicts these steps.  

In order to avoid the members to be placed too close to each other or to overlap, we compute the offsets  
\begin{equation*}
d^{(+)}\left(\beta(u_1), \gamma(u_1), w_m\right) \quad\text{and}\quad d^{(-)}\left(\beta(u_1), \gamma(u_1), w_m\right)
\end{equation*}
which give the minimum distance between a member and its preceding and subsequent neighbors. 
The angles $\beta(u_1)$ and $\gamma(u_1)$ are the enclosed angles between a member and the boundaries, and $w_m$ is the member width (cf.~{Figure~\ref{fig:alpha}}). 

If members are too dense, we prioritize them using the absolute value of curvature of $\distf_u(s)$. 
The assumption behind this choice is inspired by the observation that the more curved $\distf_u$ locally is, the more distinct surface features the corresponding geodesic captures. 

If members are too sparse, we add new members  in the gaps, which fulfill the restrictions imposed by $d^{(+)}$ and $d^{(-)}$.
After adding them, we minimize the sum of the squared distances to existing members in order to achieve a more equal distribution.

Note that the same procedure is applied to $\mathcal{D}_v$ and $\tod{\mathcal{D}}_v$ to obtain the function $\distf_v$ and the members of the ($h, \tod{h}$) family. 
 
\begin{figure}
	\centering
	\includegraphics[width=1\columnwidth]{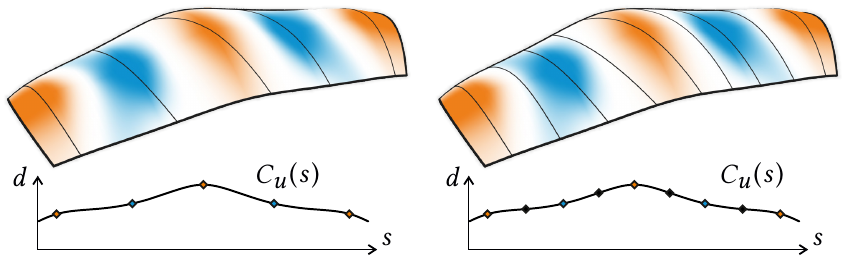}
	\caption{One iteration of the member placement procedure. Left: members placed based on geometric features. Right: additional members placed in the gaps and distributed without affecting the initial members. Bottom row depicts the $\distf$-function with indicated members (cf.~Section~\ref{sec:members}).} 
	\label{fig:densify}
\end{figure}

\subsection{Notches}\label{sec:notches}

Deploying the planar grid with rotational-only connections delivers an approximation of the surface patch $\Psur$, but the centerlines of the physical lamellas cannot become geodesics on $\Psur$.
The reason is that they are held back by their fixed intersections with inner members of the other family. This restriction is a consequence of the grid criteria~\ref{GC1} and~\ref{GC2}. 
Note that as shown by Lagally~\shortcite{Lagally}, an arbitrary geodesic grid cannot be planarized in general. 

To address this issue,  we introduce \emph{sliding notches} at the connections of inner members. 
These notches provide two translational degrees of freedom at each connection, enabling the respective members $g_i$ and $h_j$ to slide by the notch lengths $\ell_{g_i}$, $\ell_{h_j}$ (cf. Figure~\ref{fig:notches}). 
We can identify unique optimal sliding directions and notch lengths from comparing the difference of the locations of the connections w.r.t. the arc length between the geodesic members $\gc,\hc$ and their planar counterparts $\tod{\gc},\tod{\hc}$.

In other words, traversing an inner member pair $(g_i(s),\tod{g_i}(\tod{s})) \in (\gc,\tod{\gc})$ along its arc length parameters $s$ and $\tod{s}$, the notch length $\ell_{g_i}$ at a particular connection is given by 
\[
\ell_{g_i} = s-\tod{s} \,. 
\]
The notch length $\ell_{h_i}$ along the $(h_i(s),\tod{h_i}(\tod{s}))$ member pair is given in an analogous way (cf. Figure~\ref{fig:notches}).

The corresponding sliding directions are given by the sign of this equation. 
If each connection slides to the end of both its notches, the centerlines of the lamellas move towards the geodesics on $\Psur$. 
Due to the extra degrees of freedom, notches enable the structure to take a lower energy state by reducing the torsion and curvature of the members.
The notches are physically realized by simply elongating the holes of the corresponding lamellas.

\begin{figure}[t]
	\centering
	\includegraphics[width = 0.49\textwidth]{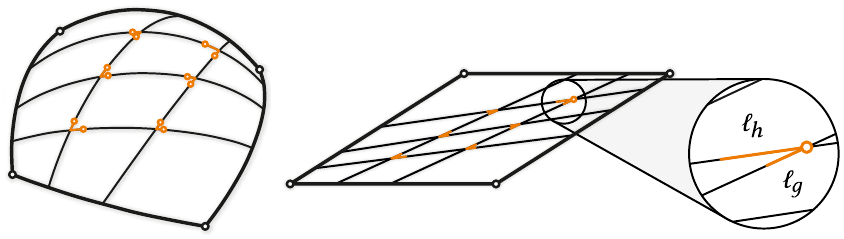}
	\caption{Left: deployment without notches, where orange dots indicate optimal connections in the spatial state. Right: Notches $\ell_{g}$, $\ell_{h}$ computed for one particular connection $q$ (cf. Section~\ref{sec:notches}).   }
	\label{fig:notches}
	\vspace{-6pt}
\end{figure}

\subsection{Anchors}\label{sec:anchors}

When changing the angle \begin{math} \tod{\alpha} \rightarrow \alpha \end{math}, an elastic grid buckles out of plane into a curved configuration. 
While the surface patch $\Psur$ has a fixed shape, the grid can deform to multiple spatial configurations, since an elastic grid for a specific surface patch is also suitable for all isometric surface patches. This is given by the fact that our grids are constructed using the intrinsic metric on $\Psur$, which is invariant to isometries. 
Isometries of a surface can be imagined by bending the surface without stretching it. 

To force the grid into the desired configuration, we introduce additional anchors which pin connections of members to fixed points on the target surface. 
We systematically introduce them on selected connections of inner members with boundary curves, such that  they push the elastic grid into a configuration in agreement with the shape of $\Psur$. 

For practical reasons, we only allow anchors on the boundaries. 
In particular, we identify points of locally extreme curvature on the boundary geodesics and filter for small extrema. The connections of members closest to these points serve as anchor locations (cf.~Fig.~\ref{fig:comparison}).

\section{Physical Simulation}\label{sec:model}

To simulate the physical behavior of the deployed grid, we use a simulation based on discrete elastic rods \cite{Bergou2010} and build upon the solution of~\cite{Vekhter2019}. We refer the reader to those papers for the details. 
Note, that the associated material frames of the rods do not need to be isotropic, which allows us also to model the exact cross sections of lamellas with a ratio of $1:10$.

A central aspect of the kinematics of elastic geodesic grids is the ability of grid members to slide at connections, denoted in the following as $q$. In general, they do not coincide with the vertices of the discretized grid members. 
To handle them, we introduce barycentric coordinates $\beta_{q}$ to describe the location of a connection on a rod-edge. 
We also take the physical thickness $t$ of the lamellas into account, which is modeled by an offset between the members $g$ and $h$ at each connection. Hence, a connection $q$ consists of two points $q_g$  and $q_h$ with an offset $t$. 
Apart from sliding, members are allowed to rotate around connections about an axis that is parallel to the cross product of the edges $q_g$ and $q_h$  lie on.

\paragraph{Simulation} 

\begin{figure}[t]
	\centering
	\includegraphics[width=1.01\columnwidth]{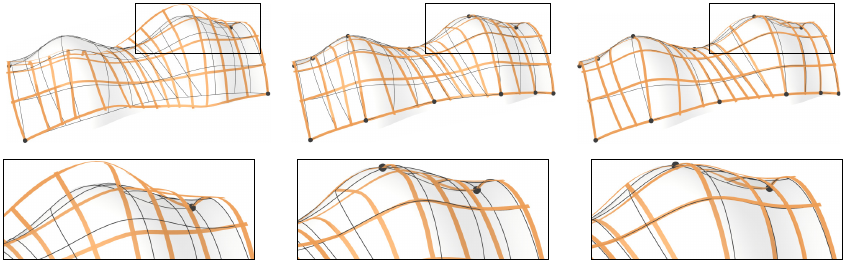}
	\caption{The influence of anchors and notches on the example \textit{Archway}. Left: Anchors at the corners are not sufficient to push the grid into the right configuration. Center: Deployed state without notches, local buckling and irregularities in smoothness can be observed. Right: Notches relax the structure to a more natural, lower energy shape (cf. Sections~\ref{sec:notches} and~\ref{sec:anchors}). }
	\label{fig:comparison}
	\vspace{-6pt}
\end{figure}

Our aim is to find the equilibrium state of the given elastic grid, which corresponds to an optimization problem of minimizing the energy functional 
\begin{equation*}
E = E_r + E_q + E_a + E_n + E_p,
\end{equation*}
where $E_r$ is the internal energy of the rods, $E_q$ is the energy of the connection constraints, $E_a$ is the energy of the anchor constraints, $E_n$ is the energy of the notch-limit constraints, and $E_p$ is an additional notch penalty term that also serves to account for friction.
We perform the simulation by minimizing the entire energy $E$ for the rod centerline points ${x}$ using a Gauss-Newton method in a similar fashion as proposed by Vekhter \etal \shortcite{Vekhter2019}. 
In Section \ref{sec:results:sim} we perform an empirical evaluation of the accuracy of the simulation by comparing it to laser-scans of the makes. 

For the sake of readability, we will define the constraint energy terms only for a single constraint each. 
$E_r$ is the sum of stretching, bending and twisting energies of each individual rod. As a full explanation of the DER formulation is out of scope for this paper, we refer the reader to the work of \cite{Bergou2010} for a detailed description of these terms.

The connection constraint energy $E_q$ is given by
\begin{equation*}
\begin{aligned}
E_q =~ &\w_{q,1} \norm{ q_{g} - q_{h} + t {m}_{g} }^2 
+ \w_{q,1} \norm{ q_{h} - q_{g} - t {m}_{h} }^2 \\
+ ~&\w_{q,2} \norm{  \angle \left( {m}_{g}, {m}_{h} \right) }^2 ,
\end{aligned}
\end{equation*}
with ${m}_{g}$ and ${m}_{h}$ denoting the material vectors of $g$ and $h$ at $q$ respectively. 
The term $t{m}$ accounts for the thickness of the rods, while $\w_{q,1}$ and $\w_{q,2}$ are the constraint weights for the position and direction terms.

The anchor constraint energy $E_a$ ensures that both the position $q$ and material vector $m$ of the given connection do not deviate from the position ${q}_a$ and material vector ${m}_a$ of the corresponding anchor. It is given by
\begin{equation*}
\begin{aligned}
E_a = \w_{a,1} \norm{ q - {q}_a }^2 + \w_{a,2} \norm{  \angle \left( {m}, {m}_a \right) }^2 ,
\end{aligned}
\end{equation*}
with $\w_{a,1}$ and $\w_{a,2}$ as weights. 
This constraint applies to the grid corners and anchors.

The notch-limit constraint energy $E_n$ ensures that the connection point remains within the bounds of the notch. They are specified by the notch length $\notchl$  and the sliding direction (cf. Section~\ref{sec:notches}):   
\begin{equation*}
\begin{aligned}
E_n = \delta^{\left(-\right)} \left( \frac{1}{10} \log \left( \beta_q - \beta^{(-)} \right) \right)^2 + \delta^{(+)} \left( \frac{1}{10} \log \left( \beta^{(+)} - \beta_q \right) \right)^2,
\end{aligned}
\end{equation*}
with $\beta^{(-)}$ and $\beta^{(+)}$ denoting the barycentric coordinates of the notch bounds on their corresponding edges. The term is only active when the connection lies on the same rod-edge as one of the notch bounds, so $\delta^{(-)} = 1$ or $\delta^{(+)} = 1$ when the connection lies on one of these edges, and $0$ otherwise.

The additional notch penalty term $E_p$ controls the movement of a connection $q$ between two adjacent edges.
If $q$ switches edges, it needs to be reprojected to the neighboring edge at the next iteration of the simulation. 
Within an iteration, $E_p$ prevents $q$ from moving too far beyond the end of the current edge:
\begin{equation*}
\begin{aligned}
E_p = \left( \mu \log \left( \epsilon + \beta_q \right) \right)^2 +  \left( \mu \log \left( \epsilon + 1 - \beta_q \right) \right)^2,
\end{aligned}
\end{equation*}
with $\epsilon$ denoting how far $q$ is allowed to move past the end of the edge  and $\mu$ acting as a weighting parameter (we choose $\epsilon = 0.0001$, $\mu =0.1$). 

Since $E_p$ is not $0$ even inside the edge, it penalizes very small sliding movements that would otherwise accumulate over many iterations. In other words, $E_p$ creates a pseudo-frictional effect, which is controlled by $\mu$. 
In a physical grid, friction creates a force acting against the sliding movement of a connection. If the driving force of the movement and the frictional force counterbalance, the movement stops. 
This situation has an analogy in our grids. A connection stops moving inside a notch if 
\begin{equation*}
\begin{aligned}
\frac{\partial E_q}{\partial {\beta_q}} +  \frac{\partial E_p}{\partial {\beta_q}}= 0
\end{aligned}
\end{equation*}
is fulfilled.
Figure~\ref{fig:friction} depicts the effects of different values for $\mu$.

\begin{figure}[t]
	\centering
	\includegraphics[width=1\columnwidth]{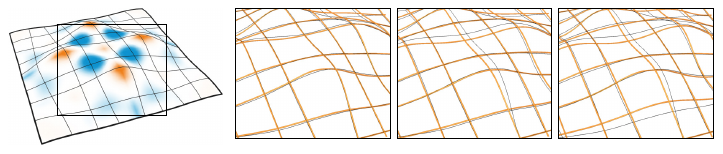}
	\caption{The effect of the weighting parameter $\mu$ in $E_p$ (from left to right): surface shaded with $K$ and geodesics;  $\mu = 0.01$, rods slide onto geodesics; $\mu = 0.1$, sliding in high $K$ areas reduced (our setting); $\mu = 1$, sliding is heavily reduced. Refer to Section \ref{sec:friction} for a further discussion on $\mu$.}
	\label{fig:friction}
\vspace{-6pt}
\end{figure}

\section{Results and Evaluation}\label{sec:results}

\subsection{Qualitative Results and Fabrication}

Using our method, we have approximated a number of surfaces which are depicted in Figures \ref{fig:results2} and \ref{fig:results1}. We used input surfaces with positive and negative Gaussian curvature regions, as well as purely elliptic and hyperbolic surfaces.

The fabricated models we present in Figure \ref{fig:results1} are made of lime wood lamellas and placed on 3d-printed supports after assembly.
To position  the notches precisely, lamellas are laser-cut from thin lime wood plates. Members are connected by simply using screws and nuts. The support structures fix the shape of the boundary members to anchors as described in Section \ref{sec:anchors} and also provide correct orientation for the lamellas by inclined contact areas.

\subsection{Evaluation}

\paragraph{Quantitative Results}
In Table \ref{tab:commands} we summarize quantitative results of our method for seven models (Figure \ref{fig:results2} and \ref{fig:results1}). The presented values $\text{RMS}_{1}$ and $\text{RMS}_{2}$ denote the root mean square distance between grid vertices and the mesh representing $\Psur$ without and with notches respectively. As can be seen, notches allow for closer proximity between the rods and $\Psur$. 
Please note that the model width, depth and height listed in Table \ref{tab:commands} are dimensionless and that we scale the model by a global factor for fabrication.

The computation time for the geometric grid generation (c.f. Section \ref{sec:geometry}) mainly depends on the mesh resolution of $\Psur$, which also determines the number of distance fields that are computed. Smoothing additionally requires the computation of several distance fields in every iteration. 
Simulation time of the deployed state of the grid with and without notches mainly depends on the number of grid vertices.

\paragraph{Evaluation of Simulation}\label{sec:results:sim}

To evaluate the agreement of the simulated results with the fabricated wooden makes, we used a state-of-the-art laser-scanning device (Metris MCA 36M7) to capture the deployed gridshell. 
To enable precise agreement of the cartesian anchor coordinates $q_a$  and the point cloud, we registered them using the ICP algorithm. 

The material properties of the wood were not determined by testing, but estimated using reference values for deciduous woods. 
Figure \ref{fig:scans} shows the results of the comparison. 
Note that the root mean square error between the point cloud and the simulated model is 0.06 cm, which is only about half the thickness of a lamella.

\begin{figure}[t]
	\centering
	\includegraphics[width=0.47\textwidth]{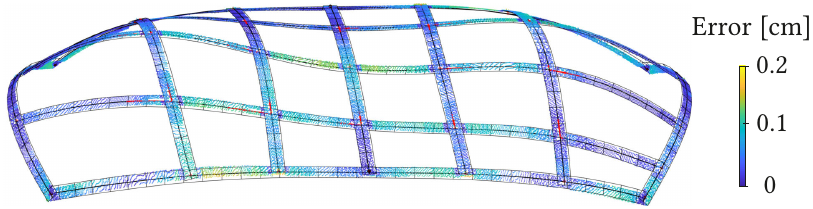}
	\caption{Comparison of the simulation result (Section~\ref{sec:model}) to a laser scan of the example \textit{Double Vault}. The figure shows the point cloud with simulation results overlayed. The notches are indicated in red. The lamellas have cross section of $0.1:1.0$ cm. The color indicates the $L^2$ distances of the points to the lamellas. The total RMS error of the comparison is $0.06$ cm. }
	\label{fig:scans}
	
	\vspace{-10pt}
\end{figure}

\begin{table}[b]
	\small
	\caption{Quantitative results of our method. We measure the root mean square error (RMS) between the member centerlines and the target mesh: $\text{RMS}_{1}$ refers to grids without notches and $\text{RMS}_{2}$ to grids with notches. Timings are in seconds,  $t_{\text{grid}}$ refers to the computation times of generating the geometric elastic grid, $t_{1}$ refers to the simulation without notches and $t_{2}$ to the simulation with notches. $|M_V|$ expresses the number of mesh vertices and $|G_V|$ the number of grid vertices. Captions refer to examples Torus Wide,  Waves Bump (Fig.~\ref{fig:results2}), and Sphere, Double Vault, Waves, Archway, and Triple Vault (Fig.~\ref{fig:results1}) respectively. Measured on an Intel~Xeon~E5-2687W~v4.}
	\label{tab:commands}
\begin{tabular}{r d{1} d{1} d{1} d{1} d{1} d{1} d{1}}
	& \multicolumn{1}{c}{T.W.} & \multicolumn{1}{c}{W.B.} & \multicolumn{1}{c}{Sph.} &  \multicolumn{1}{c}{D.V.}& \multicolumn{1}{c}{W.}   &\multicolumn{1}{c}{A.w.} & \multicolumn{1}{c}{T.V.}\\
	\midrule
	width 	& 100.0 & 100.0 & 100.0 & 100.0 & 100.0 & 100.0 & 100.0 \\
	depth 	  &  61.9 & 100.0&	100.0 & 	51.7 &  65.5  & 58.0 & 42.8\\
	height 	 &	27.2 &  12.7  &	29.9  &  14.6 &  15.1 & 20.7& 16.3\\[0.1cm]
	
	$|M_V|$   &    \multicolumn{1}{r}{2122}&    \multicolumn{1}{r}{3385}  &    \multicolumn{1}{r}{1083} &	  \multicolumn{1}{r}{571}& \multicolumn{1}{r}{1929}	& \multicolumn{1}{r}{975} &    \multicolumn{1}{r}{1322} \\		
	$|G_V|$   &    \multicolumn{1}{r}{767}	&    \multicolumn{1}{r}{388}  &   \multicolumn{1}{r}{414} &   	  \multicolumn{1}{r}{300}& \multicolumn{1}{r}{328}	 &   \multicolumn{1}{r}{625} &  \multicolumn{1}{r}{494} \\[0.1cm]

	\midrule
	$t_{\text{smooth}}$ & \multicolumn{1}{c}{$-$}  & \multicolumn{1}{r}{31.63}  & \multicolumn{1}{c}{$-$} & \multicolumn{1}{c}{$-$}  & \multicolumn{1}{r}{ 10.22}&\multicolumn{1}{r}{4.14} & \multicolumn{1}{c}{$-$}\\

	$t_{\text{grid}}$ & \multicolumn{1}{r}{5.33} & \multicolumn{1}{r}{5.62}  & \multicolumn{1}{r}{1.29}&	\multicolumn{1}{r}{0.68} & \multicolumn{1}{r}{2.10} &	\multicolumn{1}{r}{1.50} & \multicolumn{1}{r}{1.67}\\
	\midrule

	$\text{RMS}_{1}$   &	\multicolumn{1}{r}{1.17}     &	\multicolumn{1}{r}{1.47}  &	\multicolumn{1}{r}{1.09}&  \multicolumn{1}{r}{0.69}  & 	\multicolumn{1}{r}{0.59}&	\multicolumn{1}{r}{0.63} &\multicolumn{1}{r}{0.69}  \\
	$\text{RMS}_{2}$	 &	\multicolumn{1}{r}{0.27}&	\multicolumn{1}{r}{0.78} &	\multicolumn{1}{r}{0.58}& 	 \multicolumn{1}{r}{0.31}  & 	\multicolumn{1}{r}{0.43}  &	\multicolumn{1}{r}{0.42} &\multicolumn{1}{r}{0.46}  \\[0.1cm]

	$t_{1}$ & \multicolumn{1}{r}{1.92} & \multicolumn{1}{r}{12.60}& \multicolumn{1}{r}{6.05}&	\multicolumn{1}{r}{2.25}  & \multicolumn{1}{r}{3.03}  &	\multicolumn{1}{r}{37.74} & \multicolumn{1}{r}{3.50} \\

	$t_{2}$  & \multicolumn{1}{r}{6.48} & \multicolumn{1}{r}{57.22}& \multicolumn{1}{r}{4.25} & \multicolumn{1}{r}{4.05} & \multicolumn{1}{r}{9.56} & 	\multicolumn{1}{r}{85.43} &\multicolumn{1}{r}{5.80} \\						
\end{tabular}
\end{table}

\subsection{Implementation}
Our grid design algorithm is implemented in \matlab, utilizing its sequential quadratic programming solver for solving the optimization Problem (\ref{eq:alpha}) using numerical gradients w.r.t. $\alpha$. We furthermore implemented the DER-simulation in C\texttt{++}, building upon the framework of \cite{Vekhter2019}. To compute the distance fields on the surface patch $\Psur$ we use the VTP algorithm by \cite{qin16}. For the computation of the geodesic paths we use the algorithm for exact geodesics between two points by \cite{surazhsky05}.

\begin{figure*}
	\centering
	\includegraphics[width=0.999\textwidth]{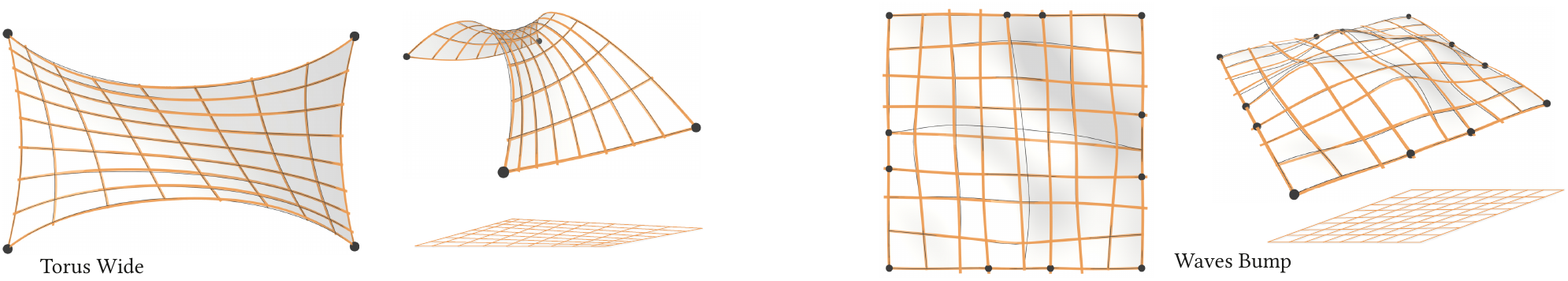} 
	\caption{Computed and simulated results without make, renderings of the simulation and the planar grid. The orange lines follow our simulation with notches. The dark lines follow the shortest geodesics on $\Psur$. }
	
	\label{fig:results2}
	\vspace{-5pt}
\end{figure*}

\section{Discussion}\label{sec:discussion}

\subsection{Geodesic Grids vs General Grids}

In order to design general grids, the paths of the surface curves need to be flexible. 
In our method, we focus on geodesic curves due to their properties, in particular allowing only the normal curvature on surfaces  (cf. Section~\ref{sec:definition}). 
The directions of the curves on the surface can only be controlled by changing the angle $\tod{\alpha}$ because of the restrictions induced by the cladding functions.
Creating an elastic geodesic grid that approximates an arbitrary curve network is therefore not possible.

As a consequence of our design choice, cross sections of fabricated members need to be rectangular with a high width to thickness ra\-tio. While this ensures easy fabrication, at the same time it poses a limitation on the design space. 
As shown by Panetta et al.~\shortcite{Panetta2019}, the shape-space of similar grid structures can be controlled by changing the profile of cross sections. 
However, when using more complicated cross sections, parts of them may buckle during deployment. 
This causes nonlinearities in stiffness parameters requiring to account for buckled cross sections. 
We avoid this necessary nontrivial update of the stiffness parameters, as the choice of our cross section minimizes these geometric second order effects.

Note that in our models, the size of the cross sections is uniform. Allowing different dimensions for every rod or even every segment would allow for an even better approximation of the surface patch.

\subsection{Representable Shapes}

Elastic geodesic grids can only approximate surfaces, that are ``clad\-dable'' by unique shortest geo\-de\-sics. If this is not the case, our smoothing algorithm ensures cladding, but surface details could be lost. 
Also the number and the density of members influences the representable shapes. If the shape is of very high frequency geometric details, it might not be representable by a too sparse network of physical members. In turn, in order to ensure fabricability, only a limited number of members is possible. This relationship is an interesting issue for future work. 

To approximate the extrinsic shape of $\Psur$, we introduce anchors on the boundaries of an elastic grid. They act as constraints on the shape of the grid and are supposed to reduce the number of possible configurations to a single one.
However, in some cases our definition of anchors is not sufficient. Imagine a high-frequency surface: fixed boundaries may not suffice to uniquely determine the direction of inner bumps.
Although we did not encounter this problem in our examples, there certainly exist surface patches that require additional anchors inside the grid to pin down its shape uniquely.

Besides this geometric view on multiple deployed configurations, they can also be looked at from an equilibrium point of view. 
If deployed and anchored correctly, a structure in equilibrium will maintain its shape. 
Further conclusions about the nature of the equilibrium would require a sensitivity analysis which could give interesting insights to the properties of elastic grids like the proneness to pop into a different configuration in a loading scenario. 

Notches allow the grid to relax into a lower energy state and increase the accuracy of the approximation.
If a grid without notches is deployed, it cannot approximate the surface patch $\Psur$, because distances between connections do not agree with the metric of $\Psur$.
The effects can be observed in local buckling of members and general deviations from $\Psur$ (cf. Figure \ref{fig:comparison}).

Finally, the current definition of distance maps is not compatible with holes in the surface, so the surface patch needs to maintain a single boundary.

\vspace{-4pt}
\subsection{Simulation}\label{sec:friction}

In our simulation, the energy term $E_p$  is not physical, nonetheless, it acts as a source of pseudo-friction. We incorporated it to speed up the convergence of sliding movements and to make the simulation more realistic. 
As $E_p$ causes connections to not fully utilize the notches, it interferes with the quality of the approximation (cf. Figure~\ref{fig:comparison}).

However, in our simulated models we registered that successively increasing $\mu$ first penalizes notches that belong to members with geodesics in areas of high $K$. Here geodesics are sensitive to imprecisions (e.g., from discretization of $\Psur$ or our numeric algorithm) and can exhibit deviations from the desired optimal path. This results in notches that are overly long.

The effects of $E_p$ penalize sliding in high $K$ regions first, which helps to trim such locally overly long notches (c.f. Figure \ref{fig:results2}, Waves Bump and Figure \ref{fig:results1}, Archway). 
Using the suggested settings, there is no significant negative effect of  $E_p$ on the quality of approximation as Table \ref{tab:commands}  and the Figures \ref{fig:results2} and \ref{fig:results1} show. 
It would be interesting to investigate a notch-penalty term that goes beyond imitating friction, but controlling the quality of the approximation via systematically reducing notch-lengths.  A further investigation into similar concepts of handling notches is an attractive topic for future work.
 
The used simulation is based on the DER formulation and therefore uses the concept of linear material elasticity. It does not account for non-linear elastic effects like plasticity or the failure of members. Since we prescribe deformations in the deployment scenario, the resulting stresses have to be kept within an acceptable range. These arising stresses are higly influenced by crosssectional sizing.

\vspace{-4pt}
\subsection{Deployment}
The deployment of an elastic grid is achieved by changing the angle $\tod{\alpha}$ and applying additional bending to guide it to the desired extrinsic shape. 
While our treatment of the deployment process is limited to the start and end configurations, without investigating intermediate states, we expect the process to be feasible if the end configuration is physically sound.
All our experiments performed in accordance with this expectation, although a proof remains future work.

While deploying our physical models, we encountered that the static friction of wood can hinder connections from sliding freely. It thereby prevents the system from moving into a configuration of lower elastic energy. This can be countered by introducing some extra energy into the system that helps to overcome friction. Also finding fabrication methods that minimize friction between members are interesting problems to explore in the future.

Our approach is intended as a form-finding tool for 2d-3d elastically deployable gridshell structures. 
Although we only validated our approach with small scale models, \cite{Panetta2019} examined the deployment of structures that use a similar deployment mechanism, but are bigger in size. Investigating how our approach can be adapted to the challenges of large scale architecture is an interesting engineering problem and a potential topic for future work.

\begin{figure*}
	\centering
	\includegraphics[width=1.0\textwidth]{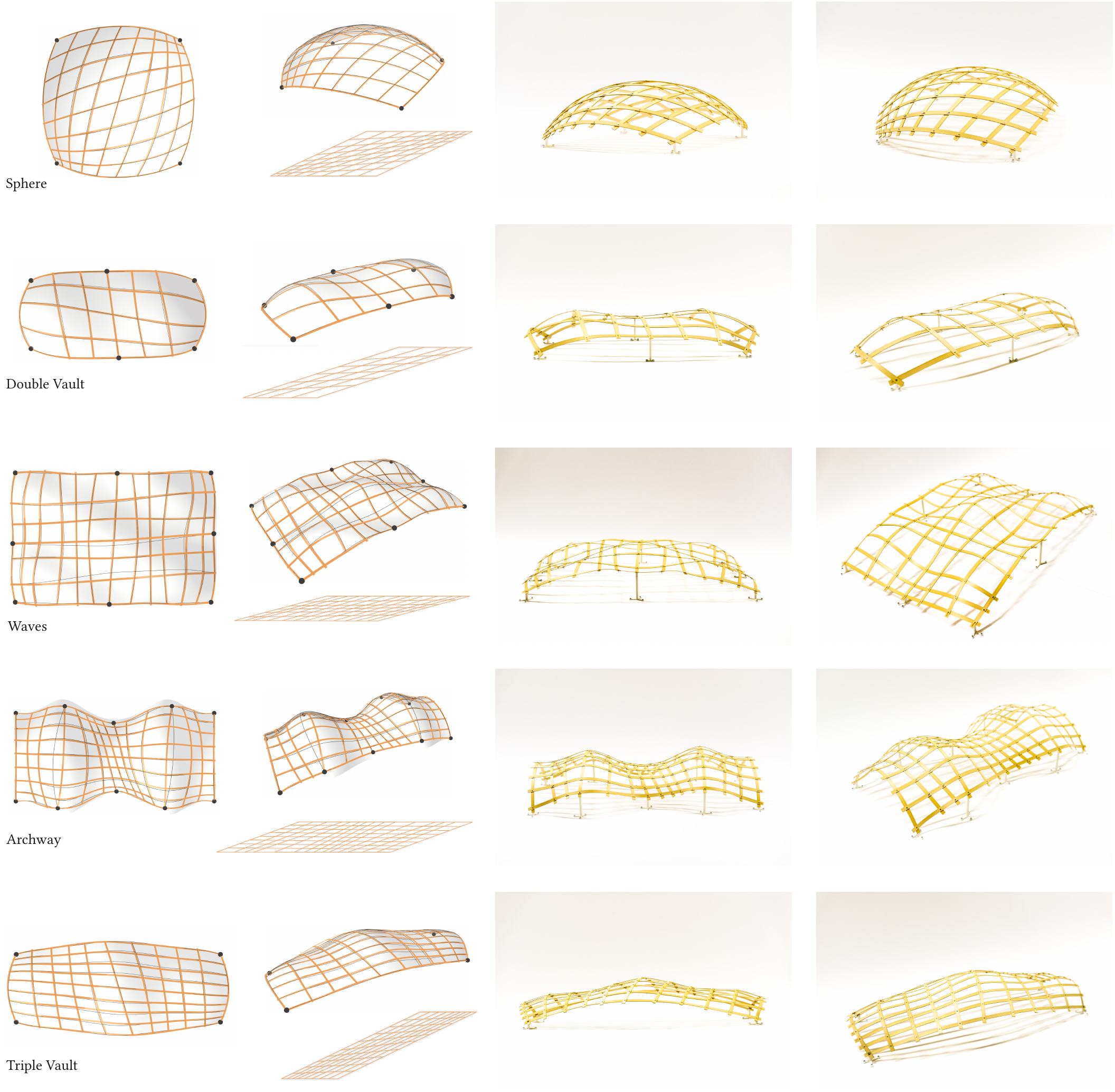} 
	\caption{Computed, simulated, and fabricated results of our method. Left: computed planar grids and renderings of the simulation. The orange strips follow our simulation with notches, the dark lines follow the shortest geodesics on $\Psur$. Right: photographs of our makes.  
	Best seen in the electronic version in closeup. }
	\label{fig:results1}
	\vspace{-6pt}
\end{figure*}

\vspace{-4pt}
\section{Conclusions}\label{sec:end}

We presented a novel approach for computational design of elastic gridshell structures that approximate smooth freeform surfaces by placing grid elements close to geodesic curves on the surface. Our method is inspired by architecture and design, and aims at simple fabrication, assembly, and most importantly at easy planar--to--spatial deployment. Moreover, it should provide an easy to handle tool for designers to create physically sound and aesthetically pleasing spatial grid structures based on the \textit{active bending} paradigm.   

Our solution is based on theoretical considerations and combines geometrical background with physical simulation. 
We have proposed a concept for the computation and simulation of such elastic grids. Additionally, we compared the results of the simulation to real fabricated grids and show that they match very well. Finally, we presented a set of examples with varying Gaussian curvature and fabricated a subset of them as wooden small-scale gridshells as a proof of our concept.

\begin{acks}
This research was mainly funded by the \grantsponsor{WWTF ICT15-082}{Vienna Science and Technology Fund (WWTF ICT15-082)}~~and  partially also by the  \grantsponsor{FWF P27972-N31}{Austrian  Science  Fund  (FWF P27972-N31)}~. 
The authors thank Florian Rist, Christian M\"uller, and Helmut Pottmann for inspiring discussions, as well as Etienne Vouga and Josh Vekhter for sharing code.   
\end{acks}

\bibliographystyle{ACM-Reference-Format}
\bibliography{papers-elasticgrids,additional-refs}


\begin{thebibliography}{50}


\ifx \showCODEN    \undefined \def \showCODEN     #1{\unskip}     \fi
\ifx \showDOI      \undefined \def \showDOI       #1{#1}\fi
\ifx \showISBNx    \undefined \def \showISBNx     #1{\unskip}     \fi
\ifx \showISBNxiii \undefined \def \showISBNxiii  #1{\unskip}     \fi
\ifx \showISSN     \undefined \def \showISSN      #1{\unskip}     \fi
\ifx \showLCCN     \undefined \def \showLCCN      #1{\unskip}     \fi
\ifx \shownote     \undefined \def \shownote      #1{#1}          \fi
\ifx \showarticletitle \undefined \def \showarticletitle #1{#1}   \fi
\ifx \showURL      \undefined \def \showURL       {\relax}        \fi
\providecommand\bibfield[2]{#2}
\providecommand\bibinfo[2]{#2}
\providecommand\natexlab[1]{#1}
\providecommand\showeprint[2][]{arXiv:#2}

\bibitem[\protect\citeauthoryear{Baek and Reis}{Baek and Reis}{2019}]%
        {Baek2019}
\bibfield{author}{\bibinfo{person}{Changyeob Baek} {and}
  \bibinfo{person}{Pedro~M. Reis}.} \bibinfo{year}{2019}\natexlab{}.
\newblock \showarticletitle{{Rigidity of hemispherical elastic gridshells under
  point load indentation}}.
\newblock \bibinfo{journal}{\emph{Journal of the Mechanics and Physics of
  Solids}}  \bibinfo{volume}{124} (\bibinfo{date}{March} \bibinfo{year}{2019}),
  \bibinfo{pages}{411--426}.
\newblock
\showISSN{0022-5096}
\urldef\tempurl%
\url{https://doi.org/10.1016/J.JMPS.2018.11.002}
\showDOI{\tempurl}


\bibitem[\protect\citeauthoryear{Bergou, Audoly, Vouga, Wardetzky, and
  Grinspun}{Bergou et~al\mbox{.}}{2010}]%
        {Bergou2010}
\bibfield{author}{\bibinfo{person}{Mikl{\'{o}}s Bergou},
  \bibinfo{person}{Basile Audoly}, \bibinfo{person}{Etienne Vouga},
  \bibinfo{person}{Max Wardetzky}, {and} \bibinfo{person}{Eitan Grinspun}.}
  \bibinfo{year}{2010}\natexlab{}.
\newblock \showarticletitle{{Discrete viscous threads}}.
\newblock \bibinfo{journal}{\emph{ACM Transactions on Graphics}}
  \bibinfo{volume}{29}, \bibinfo{number}{4} (\bibinfo{date}{July}
  \bibinfo{year}{2010}), \bibinfo{pages}{1}.
\newblock
\showISBNx{9781450301121}
\showISSN{07300301}
\urldef\tempurl%
\url{https://doi.org/10.1145/1778765.1778853}
\showDOI{\tempurl}


\bibitem[\protect\citeauthoryear{Bergou, Wardetzky, Robinson, Audoly, and
  Grinspun}{Bergou et~al\mbox{.}}{2008}]%
        {Bergou2008}
\bibfield{author}{\bibinfo{person}{Mikl{\'{o}}s Bergou}, \bibinfo{person}{Max
  Wardetzky}, \bibinfo{person}{Stephen Robinson}, \bibinfo{person}{Basile
  Audoly}, {and} \bibinfo{person}{Eitan Grinspun}.}
  \bibinfo{year}{2008}\natexlab{}.
\newblock \showarticletitle{{Discrete elastic rods}}.
\newblock \bibinfo{journal}{\emph{ACM Transactions on Graphics}}
  \bibinfo{volume}{27}, \bibinfo{number}{3} (\bibinfo{date}{Aug.}
  \bibinfo{year}{2008}), \bibinfo{pages}{1}.
\newblock
\showISBNx{9781450302104}
\showISSN{07300301}
\urldef\tempurl%
\url{https://doi.org/10.1145/1360612.1360662}
\showDOI{\tempurl}


\bibitem[\protect\citeauthoryear{Bermano, Funkhouser, and Rusinkiewicz}{Bermano
  et~al\mbox{.}}{2017}]%
        {Bermano2017}
\bibfield{author}{\bibinfo{person}{Amit~H. Bermano}, \bibinfo{person}{Thomas
  Funkhouser}, {and} \bibinfo{person}{Szymon Rusinkiewicz}.}
  \bibinfo{year}{2017}\natexlab{}.
\newblock \showarticletitle{{State of the Art in Methods and Representations
  for Fabrication-Aware Design}}.
\newblock \bibinfo{journal}{\emph{Computer Graphics Forum}}
  \bibinfo{volume}{36}, \bibinfo{number}{2} (\bibinfo{date}{May}
  \bibinfo{year}{2017}), \bibinfo{pages}{509--535}.
\newblock
\showISSN{01677055}
\urldef\tempurl%
\url{https://doi.org/10.1111/cgf.13146}
\showDOI{\tempurl}


\bibitem[\protect\citeauthoryear{Chen, Sitthi-amorn, Lan, and Matusik}{Chen
  et~al\mbox{.}}{2013}]%
        {Chen2013}
\bibfield{author}{\bibinfo{person}{Desai Chen}, \bibinfo{person}{Pitchaya
  Sitthi-amorn}, \bibinfo{person}{Justin~T. Lan}, {and}
  \bibinfo{person}{Wojciech Matusik}.} \bibinfo{year}{2013}\natexlab{}.
\newblock \showarticletitle{{Computing and Fabricating Multiplanar Models}}.
\newblock \bibinfo{journal}{\emph{Computer Graphics Forum}}
  \bibinfo{volume}{32}, \bibinfo{number}{2pt3} (\bibinfo{date}{May}
  \bibinfo{year}{2013}), \bibinfo{pages}{305--315}.
\newblock
\showISSN{01677055}
\urldef\tempurl%
\url{https://doi.org/10.1111/cgf.12050}
\showDOI{\tempurl}


\bibitem[\protect\citeauthoryear{Desbrun, Meyer, Schr{\"{o}}der, and
  Barr}{Desbrun et~al\mbox{.}}{1999}]%
        {Desbrun1999a}
\bibfield{author}{\bibinfo{person}{Mathieu Desbrun}, \bibinfo{person}{Mark
  Meyer}, \bibinfo{person}{Peter Schr{\"{o}}der}, {and}
  \bibinfo{person}{Alan~H. Barr}.} \bibinfo{year}{1999}\natexlab{}.
\newblock \showarticletitle{{Implicit fairing of irregular meshes using
  diffusion and curvature flow}}. In \bibinfo{booktitle}{\emph{Proceedings of
  the 26th annual conference on Computer graphics and interactive techniques -
  SIGGRAPH '99}}. \bibinfo{publisher}{ACM Press}, \bibinfo{address}{New York,
  New York, USA}, \bibinfo{pages}{317--324}.
\newblock
\showISBNx{0201485605}
\urldef\tempurl%
\url{https://doi.org/10.1145/311535.311576}
\showDOI{\tempurl}


\bibitem[\protect\citeauthoryear{Deuss, Panozzo, Whiting, Liu, Block,
  Sorkine-Hornung, and Pauly}{Deuss et~al\mbox{.}}{2014}]%
        {Deuss2014}
\bibfield{author}{\bibinfo{person}{Mario Deuss}, \bibinfo{person}{Daniele
  Panozzo}, \bibinfo{person}{Emily Whiting}, \bibinfo{person}{Yang Liu},
  \bibinfo{person}{Philippe Block}, \bibinfo{person}{Olga Sorkine-Hornung},
  {and} \bibinfo{person}{Mark Pauly}.} \bibinfo{year}{2014}\natexlab{}.
\newblock \showarticletitle{{Assembling self-supporting structures}}.
\newblock \bibinfo{journal}{\emph{ACM Transactions on Graphics}}
  \bibinfo{volume}{33}, \bibinfo{number}{6} (\bibinfo{date}{Nov.}
  \bibinfo{year}{2014}), \bibinfo{pages}{1--10}.
\newblock
\showISSN{07300301}
\urldef\tempurl%
\url{https://doi.org/10.1145/2661229.2661266}
\showDOI{\tempurl}


\bibitem[\protect\citeauthoryear{do~Carmo}{do~Carmo}{1992}]%
        {Carmo1992}
\bibfield{author}{\bibinfo{person}{Manfredo do Carmo}.}
  \bibinfo{year}{1992}\natexlab{}.
\newblock \bibinfo{booktitle}{\emph{Riemannian Geometry}}.
\newblock \bibinfo{publisher}{Birkh{\"a}user}.
\newblock
\showISBNx{9783764334901}
\showLCCN{91037377}
\urldef\tempurl%
\url{https://www.springer.com/gp/book/9780817634902}
\showURL{%
\tempurl}


\bibitem[\protect\citeauthoryear{Dudte, Vouga, Tachi, and Mahadevan}{Dudte
  et~al\mbox{.}}{2016}]%
        {Dudte2016}
\bibfield{author}{\bibinfo{person}{Levi~H. Dudte}, \bibinfo{person}{Etienne
  Vouga}, \bibinfo{person}{Tomohiro Tachi}, {and} \bibinfo{person}{L.
  Mahadevan}.} \bibinfo{year}{2016}\natexlab{}.
\newblock \showarticletitle{{Programming curvature using
  origami tessellations}}.
\newblock \bibinfo{journal}{\emph{Nature Materials}} \bibinfo{volume}{15},
  \bibinfo{number}{5} (\bibinfo{date}{May} \bibinfo{year}{2016}),
  \bibinfo{pages}{583--588}.
\newblock
\showISSN{1476-1122}
\urldef\tempurl%
\url{https://doi.org/10.1038/nmat4540}
\showDOI{\tempurl}


\bibitem[\protect\citeauthoryear{Eigensatz, Kilian, Schiftner, Mitra, Pottmann,
  and Pauly}{Eigensatz et~al\mbox{.}}{2010}]%
        {Eigensatz2010b}
\bibfield{author}{\bibinfo{person}{Michael Eigensatz}, \bibinfo{person}{Martin
  Kilian}, \bibinfo{person}{Alexander Schiftner}, \bibinfo{person}{Niloy~J.
  Mitra}, \bibinfo{person}{Helmut Pottmann}, {and} \bibinfo{person}{Mark
  Pauly}.} \bibinfo{year}{2010}\natexlab{}.
\newblock \showarticletitle{{Paneling architectural freeform surfaces}}.
\newblock \bibinfo{journal}{\emph{ACM Transactions on Graphics}}
  \bibinfo{volume}{29}, \bibinfo{number}{4} (\bibinfo{date}{July}
  \bibinfo{year}{2010}), \bibinfo{pages}{1}.
\newblock
\showISSN{07300301}
\urldef\tempurl%
\url{https://doi.org/10.1145/1778765.1778782}
\showDOI{\tempurl}


\bibitem[\protect\citeauthoryear{Garg, Sageman-Furnas, Deng, Yue, Grinspun,
  Pauly, and Wardetzky}{Garg et~al\mbox{.}}{2014}]%
        {Garg2014}
\bibfield{author}{\bibinfo{person}{Akash Garg}, \bibinfo{person}{Andrew~O.
  Sageman-Furnas}, \bibinfo{person}{Bailin Deng}, \bibinfo{person}{Yonghao
  Yue}, \bibinfo{person}{Eitan Grinspun}, \bibinfo{person}{Mark Pauly}, {and}
  \bibinfo{person}{Max Wardetzky}.} \bibinfo{year}{2014}\natexlab{}.
\newblock \showarticletitle{{Wire mesh design}}.
\newblock \bibinfo{journal}{\emph{ACM Transactions on Graphics}}
  \bibinfo{volume}{33}, \bibinfo{number}{4} (\bibinfo{date}{July}
  \bibinfo{year}{2014}), \bibinfo{pages}{1--12}.
\newblock
\showISSN{07300301}
\urldef\tempurl%
\url{https://doi.org/10.1145/2601097.2601106}
\showDOI{\tempurl}


\bibitem[\protect\citeauthoryear{Gengnagel, Lienhard, Alpermann, Gengnagel, and
  Knippers}{Gengnagel et~al\mbox{.}}{2013}]%
        {gengnagel2013active}
\bibfield{author}{\bibinfo{person}{Christoph Gengnagel},
  \bibinfo{person}{Julian Lienhard}, \bibinfo{person}{Holger Alpermann},
  \bibinfo{person}{Christoph Gengnagel}, {and} \bibinfo{person}{Jan Knippers}.}
  \bibinfo{year}{2013}\natexlab{}.
\newblock \showarticletitle{{Active bending, a review on structures where
  bending is used as a self-formation process}}.
\newblock \bibinfo{journal}{\emph{International Journal of Space Structures}}
  \bibinfo{volume}{28}, \bibinfo{number}{3-4} (\bibinfo{year}{2013}),
  \bibinfo{pages}{187--196}.
\newblock


\bibitem[\protect\citeauthoryear{Guseinov, Miguel, and Bickel}{Guseinov
  et~al\mbox{.}}{2017}]%
        {Guseinov2017}
\bibfield{author}{\bibinfo{person}{Ruslan Guseinov}, \bibinfo{person}{Eder
  Miguel}, {and} \bibinfo{person}{Bernd Bickel}.}
  \bibinfo{year}{2017}\natexlab{}.
\newblock \showarticletitle{{CurveUps}}.
\newblock \bibinfo{journal}{\emph{ACM Transactions on Graphics}}
  \bibinfo{volume}{36}, \bibinfo{number}{4} (\bibinfo{date}{July}
  \bibinfo{year}{2017}), \bibinfo{pages}{1--12}.
\newblock
\showISSN{07300301}
\urldef\tempurl%
\url{https://doi.org/10.1145/3072959.3073709}
\showDOI{\tempurl}


\bibitem[\protect\citeauthoryear{Happold and Liddell}{Happold and
  Liddell}{1975}]%
        {Happold1975}
\bibfield{author}{\bibinfo{person}{Edmund Happold} {and} \bibinfo{person}{Ian
  Liddell}.} \bibinfo{year}{1975}\natexlab{}.
\newblock \showarticletitle{{Timber Lattice Roof for the Mannheim
  Bundesgartenschau}}.
\newblock \bibinfo{journal}{\emph{The Structural Engineer}}
  \bibinfo{volume}{53}, \bibinfo{number}{3} (\bibinfo{year}{1975}).
\newblock


\bibitem[\protect\citeauthoryear{Kilian, Fl{\"{o}}ry, Chen, Mitra, Sheffer, and
  Pottmann}{Kilian et~al\mbox{.}}{2008}]%
        {Kilian2008}
\bibfield{author}{\bibinfo{person}{Martin Kilian}, \bibinfo{person}{Simon
  Fl{\"{o}}ry}, \bibinfo{person}{Zhonggui Chen}, \bibinfo{person}{Niloy~J.
  Mitra}, \bibinfo{person}{Alla Sheffer}, {and} \bibinfo{person}{Helmut
  Pottmann}.} \bibinfo{year}{2008}\natexlab{}.
\newblock \showarticletitle{{Curved folding}}.
\newblock \bibinfo{journal}{\emph{ACM Transactions on Graphics}}
  \bibinfo{volume}{27}, \bibinfo{number}{3} (\bibinfo{date}{Aug.}
  \bibinfo{year}{2008}), \bibinfo{pages}{1}.
\newblock
\showISBNx{9781450301121}
\showISSN{07300301}
\urldef\tempurl%
\url{https://doi.org/10.1145/1360612.1360674}
\showDOI{\tempurl}


\bibitem[\protect\citeauthoryear{Kilian, Monszpart, Mitra, Kilian, Monszpart,
  and Mitra}{Kilian et~al\mbox{.}}{2017a}]%
        {Kilian2017}
\bibfield{author}{\bibinfo{person}{Martin Kilian}, \bibinfo{person}{Aron
  Monszpart}, \bibinfo{person}{Niloy~J. Mitra}, \bibinfo{person}{Martin
  Kilian}, \bibinfo{person}{Aron Monszpart}, {and} \bibinfo{person}{Niloy~J.
  Mitra}.} \bibinfo{year}{2017}\natexlab{a}.
\newblock \showarticletitle{{String Actuated Curved Folded Surfaces}}.
\newblock \bibinfo{journal}{\emph{ACM Transactions on Graphics}}
  \bibinfo{volume}{36}, \bibinfo{number}{3} (\bibinfo{date}{May}
  \bibinfo{year}{2017}), \bibinfo{pages}{1--13}.
\newblock
\showISSN{07300301}
\urldef\tempurl%
\url{https://doi.org/10.1145/3015460}
\showDOI{\tempurl}


\bibitem[\protect\citeauthoryear{Kilian, Pellis, Wallner, and Pottmann}{Kilian
  et~al\mbox{.}}{2017b}]%
        {Kilian2017a}
\bibfield{author}{\bibinfo{person}{Martin Kilian}, \bibinfo{person}{Davide
  Pellis}, \bibinfo{person}{Johannes Wallner}, {and} \bibinfo{person}{Helmut
  Pottmann}.} \bibinfo{year}{2017}\natexlab{b}.
\newblock \showarticletitle{{Material-minimizing forms and structures}}.
\newblock \bibinfo{journal}{\emph{ACM Transactions on Graphics}}
  \bibinfo{volume}{36}, \bibinfo{number}{6} (\bibinfo{date}{Nov.}
  \bibinfo{year}{2017}), \bibinfo{pages}{1--12}.
\newblock
\showISSN{07300301}
\urldef\tempurl%
\url{https://doi.org/10.1145/3130800.3130827}
\showDOI{\tempurl}


\bibitem[\protect\citeauthoryear{Konakovi{\'{c}}, Crane, Deng, Bouaziz, Piker,
  and Pauly}{Konakovi{\'{c}} et~al\mbox{.}}{2016}]%
        {Konakovic2016}
\bibfield{author}{\bibinfo{person}{Mina Konakovi{\'{c}}},
  \bibinfo{person}{Keenan Crane}, \bibinfo{person}{Bailin Deng},
  \bibinfo{person}{Sofien Bouaziz}, \bibinfo{person}{Daniel Piker}, {and}
  \bibinfo{person}{Mark Pauly}.} \bibinfo{year}{2016}\natexlab{}.
\newblock \showarticletitle{{Beyond developable}}.
\newblock \bibinfo{journal}{\emph{ACM Transactions on Graphics}}
  \bibinfo{volume}{35}, \bibinfo{number}{4} (\bibinfo{date}{July}
  \bibinfo{year}{2016}), \bibinfo{pages}{1--11}.
\newblock
\showISSN{07300301}
\urldef\tempurl%
\url{https://doi.org/10.1145/2897824.2925944}
\showDOI{\tempurl}


\bibitem[\protect\citeauthoryear{Konakovi{\'{c}}-Lukovi{\'{c}}, Panetta, Crane,
  and Pauly}{Konakovi{\'{c}}-Lukovi{\'{c}} et~al\mbox{.}}{2018}]%
        {Konakovic-Lukovic2018}
\bibfield{author}{\bibinfo{person}{Mina Konakovi{\'{c}}-Lukovi{\'{c}}},
  \bibinfo{person}{Julian Panetta}, \bibinfo{person}{Keenan Crane}, {and}
  \bibinfo{person}{Mark Pauly}.} \bibinfo{year}{2018}\natexlab{}.
\newblock \showarticletitle{{Rapid deployment of curved surfaces via
  programmable auxetics}}.
\newblock \bibinfo{journal}{\emph{ACM Transactions on Graphics}}
  \bibinfo{volume}{37}, \bibinfo{number}{4} (\bibinfo{date}{July}
  \bibinfo{year}{2018}), \bibinfo{pages}{1--13}.
\newblock
\showISSN{07300301}
\urldef\tempurl%
\url{https://doi.org/10.1145/3197517.3201373}
\showDOI{\tempurl}


\bibitem[\protect\citeauthoryear{Lagally}{Lagally}{1910}]%
        {Lagally}
\bibfield{author}{\bibinfo{person}{Max Lagally}.}
  \bibinfo{year}{1910}\natexlab{}.
\newblock \showarticletitle{{Über die Verbiegung geodätischer Netze}}.
\newblock \bibinfo{series}{Sitzungsbericht der Bayerischen Akademie der
  Wissenschaften}, Vol.~\bibinfo{volume}{1910,10}.
  \bibinfo{publisher}{Verl.d.K.B.Akad.d.Wiss.}, \bibinfo{address}{München}.
\newblock
\urldef\tempurl%
\url{http://publikationen.badw.de/de/003396114}
\showURL{%
\tempurl}


\bibitem[\protect\citeauthoryear{Lienhard, Alpermann, Gengnagel, and
  Knippers}{Lienhard et~al\mbox{.}}{2013}]%
        {Lienhard2013}
\bibfield{author}{\bibinfo{person}{Julian Lienhard}, \bibinfo{person}{Holger
  Alpermann}, \bibinfo{person}{Christoph Gengnagel}, {and} \bibinfo{person}{Jan
  Knippers}.} \bibinfo{year}{2013}\natexlab{}.
\newblock \showarticletitle{{Active Bending, a Review on Structures where
  Bending is Used as a Self-Formation Process}}.
\newblock \bibinfo{journal}{\emph{International Journal of Space Structures}}
  \bibinfo{volume}{28}, \bibinfo{number}{3-4} (\bibinfo{date}{Sept.}
  \bibinfo{year}{2013}), \bibinfo{pages}{187--196}.
\newblock
\showISSN{0266-3511}
\urldef\tempurl%
\url{https://doi.org/10.1260/0266-3511.28.3-4.187}
\showDOI{\tempurl}


\bibitem[\protect\citeauthoryear{Lienhard and Gengnagel}{Lienhard and
  Gengnagel}{2018}]%
        {Lienhard2018}
\bibfield{author}{\bibinfo{person}{Julian Lienhard} {and}
  \bibinfo{person}{Christoph Gengnagel}.} \bibinfo{year}{2018}\natexlab{}.
\newblock \showarticletitle{{Recent developments in bending-active
  structures}}. In \bibinfo{booktitle}{\emph{Creativity in Structural Design,
  annual Symposium of the IASS – International Association for Shell and
  Spatial Structures}}. \bibinfo{address}{Boston}.
\newblock


\bibitem[\protect\citeauthoryear{Malomo, P{\'{e}}rez, Iarussi, Pietroni,
  Miguel, Cignoni, and Bickel}{Malomo et~al\mbox{.}}{2018}]%
        {Malomo2018a}
\bibfield{author}{\bibinfo{person}{Luigi Malomo}, \bibinfo{person}{Jes{\'{u}}s
  P{\'{e}}rez}, \bibinfo{person}{Emmanuel Iarussi}, \bibinfo{person}{Nico
  Pietroni}, \bibinfo{person}{Eder Miguel}, \bibinfo{person}{Paolo Cignoni},
  {and} \bibinfo{person}{Bernd Bickel}.} \bibinfo{year}{2018}\natexlab{}.
\newblock \showarticletitle{{FlexMaps}}.
\newblock \bibinfo{journal}{\emph{ACM Transactions on Graphics}}
  \bibinfo{volume}{37}, \bibinfo{number}{6} (\bibinfo{date}{Dec.}
  \bibinfo{year}{2018}), \bibinfo{pages}{1--14}.
\newblock
\showISSN{07300301}
\urldef\tempurl%
\url{https://doi.org/10.1145/3272127.3275076}
\showDOI{\tempurl}


\bibitem[\protect\citeauthoryear{Massarwi, Gotsman, and Elber}{Massarwi
  et~al\mbox{.}}{2007}]%
        {Massarwi2007a}
\bibfield{author}{\bibinfo{person}{Fady Massarwi}, \bibinfo{person}{Craig
  Gotsman}, {and} \bibinfo{person}{Gershon Elber}.}
  \bibinfo{year}{2007}\natexlab{}.
\newblock \showarticletitle{{Papercraft Models using Generalized Cylinders}}.
  In \bibinfo{booktitle}{\emph{15th Pacific Conference on Computer Graphics and
  Applications (PG'07)}}. \bibinfo{publisher}{IEEE}, \bibinfo{pages}{148--157}.
\newblock
\showISBNx{0-7695-3009-5}
\urldef\tempurl%
\url{https://doi.org/10.1109/PG.2007.16}
\showDOI{\tempurl}


\bibitem[\protect\citeauthoryear{Miguel, Lepoutre, and Bickel}{Miguel
  et~al\mbox{.}}{2016}]%
        {Miguel2016}
\bibfield{author}{\bibinfo{person}{Eder Miguel}, \bibinfo{person}{Mathias
  Lepoutre}, {and} \bibinfo{person}{Bernd Bickel}.}
  \bibinfo{year}{2016}\natexlab{}.
\newblock \showarticletitle{{Computational design of stable planar-rod
  structures}}.
\newblock \bibinfo{journal}{\emph{ACM Transactions on Graphics}}
  \bibinfo{volume}{35}, \bibinfo{number}{4} (\bibinfo{date}{July}
  \bibinfo{year}{2016}), \bibinfo{pages}{1--11}.
\newblock
\showISSN{07300301}
\urldef\tempurl%
\url{https://doi.org/10.1145/2897824.2925978}
\showDOI{\tempurl}


\bibitem[\protect\citeauthoryear{Mitani and Suzuki}{Mitani and Suzuki}{2004}]%
        {Mitani2004b}
\bibfield{author}{\bibinfo{person}{Jun Mitani} {and} \bibinfo{person}{Hiromasa
  Suzuki}.} \bibinfo{year}{2004}\natexlab{}.
\newblock \showarticletitle{{Making papercraft toys from meshes using
  strip-based approximate unfolding}}. In \bibinfo{booktitle}{\emph{ACM
  SIGGRAPH 2004 Papers on - SIGGRAPH '04}}, Vol.~\bibinfo{volume}{23}.
  \bibinfo{publisher}{ACM Press}, \bibinfo{address}{New York, New York, USA},
  \bibinfo{pages}{259}.
\newblock
\showISSN{0730-0301}
\urldef\tempurl%
\url{https://doi.org/10.1145/1186562.1015711}
\showDOI{\tempurl}


\bibitem[\protect\citeauthoryear{Panetta, Konakovi{\'{c}}-Lukovi{\'{c}},
  Isvoranu, Bouleau, and Pauly}{Panetta et~al\mbox{.}}{2019}]%
        {Panetta2019}
\bibfield{author}{\bibinfo{person}{Julian Panetta}, \bibinfo{person}{Mina
  Konakovi{\'{c}}-Lukovi{\'{c}}}, \bibinfo{person}{Florin Isvoranu},
  \bibinfo{person}{Etienne Bouleau}, {and} \bibinfo{person}{Mark Pauly}.}
  \bibinfo{year}{2019}\natexlab{}.
\newblock \showarticletitle{{X-Shells: a new class of deployable beam
  structures}}.
\newblock \bibinfo{journal}{\emph{ACM Transactions on Graphics}}
  \bibinfo{volume}{38}, \bibinfo{number}{4} (\bibinfo{date}{July}
  \bibinfo{year}{2019}), \bibinfo{pages}{1--15}.
\newblock
\showISSN{0730-0301}
\urldef\tempurl%
\url{https://doi.org/10.1145/3306346.3323040}
\showDOI{\tempurl}


\bibitem[\protect\citeauthoryear{Panozzo, Block, and Sorkine-Hornung}{Panozzo
  et~al\mbox{.}}{2013}]%
        {Panozzo2013a}
\bibfield{author}{\bibinfo{person}{Daniele Panozzo}, \bibinfo{person}{Philippe
  Block}, {and} \bibinfo{person}{Olga Sorkine-Hornung}.}
  \bibinfo{year}{2013}\natexlab{}.
\newblock \showarticletitle{{Designing unreinforced masonry models}}.
\newblock \bibinfo{journal}{\emph{ACM Transactions on Graphics}}
  \bibinfo{volume}{32}, \bibinfo{number}{4} (\bibinfo{date}{July}
  \bibinfo{year}{2013}), \bibinfo{pages}{1}.
\newblock
\showISSN{07300301}
\urldef\tempurl%
\url{https://doi.org/10.1145/2461912.2461958}
\showDOI{\tempurl}


\bibitem[\protect\citeauthoryear{P{\'{e}}rez, Otaduy, and
  Thomaszewski}{P{\'{e}}rez et~al\mbox{.}}{2017}]%
        {Perez2017a}
\bibfield{author}{\bibinfo{person}{Jes{\'{u}}s P{\'{e}}rez},
  \bibinfo{person}{Miguel~A. Otaduy}, {and} \bibinfo{person}{Bernhard
  Thomaszewski}.} \bibinfo{year}{2017}\natexlab{}.
\newblock \showarticletitle{{Computational design and automated fabrication of
  kirchhoff-plateau surfaces}}.
\newblock \bibinfo{journal}{\emph{ACM Transactions on Graphics}}
  \bibinfo{volume}{36}, \bibinfo{number}{4} (\bibinfo{date}{July}
  \bibinfo{year}{2017}), \bibinfo{pages}{1--12}.
\newblock
\showISSN{07300301}
\urldef\tempurl%
\url{https://doi.org/10.1145/3072959.3073695}
\showDOI{\tempurl}


\bibitem[\protect\citeauthoryear{P{\'{e}}rez, Thomaszewski, Coros, Bickel,
  Canabal, Sumner, and Otaduy}{P{\'{e}}rez et~al\mbox{.}}{2015}]%
        {Perez2015}
\bibfield{author}{\bibinfo{person}{Jes{\'{u}}s P{\'{e}}rez},
  \bibinfo{person}{Bernhard Thomaszewski}, \bibinfo{person}{Stelian Coros},
  \bibinfo{person}{Bernd Bickel}, \bibinfo{person}{Jos{\'{e}}~A. Canabal},
  \bibinfo{person}{Robert Sumner}, {and} \bibinfo{person}{Miguel~A. Otaduy}.}
  \bibinfo{year}{2015}\natexlab{}.
\newblock \showarticletitle{{Design and fabrication of flexible rod meshes}}.
\newblock \bibinfo{journal}{\emph{ACM Transactions on Graphics}}
  \bibinfo{volume}{34}, \bibinfo{number}{4} (\bibinfo{date}{July}
  \bibinfo{year}{2015}), \bibinfo{pages}{138:1--138:12}.
\newblock
\showISSN{07300301}
\urldef\tempurl%
\url{https://doi.org/10.1145/2766998}
\showDOI{\tempurl}


\bibitem[\protect\citeauthoryear{Pietroni, Tarini, Vaxman, Panozzo, and
  Cignoni}{Pietroni et~al\mbox{.}}{2017}]%
        {Pietroni2017}
\bibfield{author}{\bibinfo{person}{Nico Pietroni}, \bibinfo{person}{Marco
  Tarini}, \bibinfo{person}{Amir Vaxman}, \bibinfo{person}{Daniele Panozzo},
  {and} \bibinfo{person}{Paolo Cignoni}.} \bibinfo{year}{2017}\natexlab{}.
\newblock \showarticletitle{{Position-based tensegrity design}}.
\newblock \bibinfo{journal}{\emph{ACM Transactions on Graphics}}
  \bibinfo{volume}{36}, \bibinfo{number}{6} (\bibinfo{date}{Nov.}
  \bibinfo{year}{2017}), \bibinfo{pages}{1--14}.
\newblock
\showISSN{0730-0301}
\urldef\tempurl%
\url{https://doi.org/10.1145/3130800.3130809}
\showDOI{\tempurl}


\bibitem[\protect\citeauthoryear{Polthier and Schmies}{Polthier and
  Schmies}{1998}]%
        {Polthier1998}
\bibfield{author}{\bibinfo{person}{Konrad Polthier} {and}
  \bibinfo{person}{Markus Schmies}.} \bibinfo{year}{1998}\natexlab{}.
\newblock \showarticletitle{{Straightest Geodesics on Polyhedral Surfaces}}.
\newblock In \bibinfo{booktitle}{\emph{Mathematical Visualization}}.
  \bibinfo{publisher}{Springer Berlin Heidelberg}, \bibinfo{address}{Berlin,
  Heidelberg}, \bibinfo{pages}{135--150}.
\newblock
\urldef\tempurl%
\url{https://doi.org/10.1007/978-3-662-03567-2_11}
\showDOI{\tempurl}


\bibitem[\protect\citeauthoryear{Pottmann, Eigensatz, Vaxman, and
  Wallner}{Pottmann et~al\mbox{.}}{2015}]%
        {Pottmann2015a}
\bibfield{author}{\bibinfo{person}{Helmut Pottmann}, \bibinfo{person}{Michael
  Eigensatz}, \bibinfo{person}{Amir Vaxman}, {and} \bibinfo{person}{Johannes
  Wallner}.} \bibinfo{year}{2015}\natexlab{}.
\newblock \showarticletitle{{Architectural geometry}}.
\newblock \bibinfo{journal}{\emph{Computers {\&} Graphics}}
  \bibinfo{volume}{47} (\bibinfo{date}{April} \bibinfo{year}{2015}),
  \bibinfo{pages}{145--164}.
\newblock
\showISSN{00978493}
\urldef\tempurl%
\url{https://doi.org/10.1016/j.cag.2014.11.002}
\showDOI{\tempurl}


\bibitem[\protect\citeauthoryear{Pottmann, Huang, Deng, Schiftner, Kilian,
  Guibas, and Wallner}{Pottmann et~al\mbox{.}}{2010}]%
        {Pottmann2010}
\bibfield{author}{\bibinfo{person}{Helmut Pottmann}, \bibinfo{person}{Qixing
  Huang}, \bibinfo{person}{Bailin Deng}, \bibinfo{person}{Alexander Schiftner},
  \bibinfo{person}{Martin Kilian}, \bibinfo{person}{Leonidas Guibas}, {and}
  \bibinfo{person}{Johannes Wallner}.} \bibinfo{year}{2010}\natexlab{}.
\newblock \showarticletitle{{Geodesic patterns}}.
\newblock \bibinfo{journal}{\emph{ACM Transactions on Graphics}}
  \bibinfo{volume}{29}, \bibinfo{number}{4} (\bibinfo{date}{July}
  \bibinfo{year}{2010}), \bibinfo{pages}{1--10}.
\newblock
\showISBNx{978-1-4503-0210-4}
\showISSN{0730-0301}
\urldef\tempurl%
\url{https://doi.org/10.1145/1778765.1778780}
\showDOI{\tempurl}


\bibitem[\protect\citeauthoryear{Qin, Han, Yu, Yu, and Zhang}{Qin
  et~al\mbox{.}}{2016}]%
        {qin16}
\bibfield{author}{\bibinfo{person}{Yipeng Qin}, \bibinfo{person}{Xiaoguang
  Han}, \bibinfo{person}{Hongchuan Yu}, \bibinfo{person}{Yizhou Yu}, {and}
  \bibinfo{person}{Jianjun Zhang}.} \bibinfo{year}{2016}\natexlab{}.
\newblock \showarticletitle{Fast and Exact Discrete Geodesic Computation Based
  on Triangle-Oriented Wavefront Propagation}.
\newblock \bibinfo{journal}{\emph{ACM Transactions on Graphics}}
  \bibinfo{volume}{35}, \bibinfo{number}{4} (\bibinfo{date}{July}
  \bibinfo{year}{2016}), \bibinfo{pages}{1--13}.
\newblock
\showISSN{0730-0301}
\urldef\tempurl%
\url{https://doi.org/10.1145/2897824.2925930}
\showDOI{\tempurl}


\bibitem[\protect\citeauthoryear{Quinn and Gengnagel}{Quinn and
  Gengnagel}{2014}]%
        {Quinn2014}
\bibfield{author}{\bibinfo{person}{Gregory~Charles Quinn} {and}
  \bibinfo{person}{Christoph Gengnagel}.} \bibinfo{year}{2014}\natexlab{}.
\newblock \showarticletitle{{A Review Of Elastic Grid Shells, Their Erection
  Methods And The Potential Use Of Pneumatic Formwork}}.
\newblock \bibinfo{journal}{\emph{WIT Transactions on The Built Environment}}
  \bibinfo{volume}{136} (\bibinfo{date}{June} \bibinfo{year}{2014}).
\newblock
\showISBNx{978-1-84564-772-8}
\showISSN{1746-4498}
\urldef\tempurl%
\url{https://doi.org/10.2495/MARAS140111}
\showDOI{\tempurl}


\bibitem[\protect\citeauthoryear{Rabinovich, Hoffmann, and
  Sorkine-Hornung}{Rabinovich et~al\mbox{.}}{2018}]%
        {Rabinovich2018}
\bibfield{author}{\bibinfo{person}{Michael Rabinovich}, \bibinfo{person}{Tim
  Hoffmann}, {and} \bibinfo{person}{Olga Sorkine-Hornung}.}
  \bibinfo{year}{2018}\natexlab{}.
\newblock \showarticletitle{{Discrete Geodesic Nets for Modeling Developable
  Surfaces}}.
\newblock \bibinfo{journal}{\emph{ACM Transactions on Graphics}}
  \bibinfo{volume}{37}, \bibinfo{number}{2} (\bibinfo{date}{Feb.}
  \bibinfo{year}{2018}), \bibinfo{pages}{1--17}.
\newblock
\showISSN{07300301}
\urldef\tempurl%
\url{https://doi.org/10.1145/3180494}
\showDOI{\tempurl}


\bibitem[\protect\citeauthoryear{Schling, Kilian, Wang, Schikore, and
  Pottmann}{Schling et~al\mbox{.}}{2018}]%
        {Schling2018}
\bibfield{author}{\bibinfo{person}{Eike Schling}, \bibinfo{person}{Martin
  Kilian}, \bibinfo{person}{Hui Wang}, \bibinfo{person}{Jonas Schikore}, {and}
  \bibinfo{person}{Helmut Pottmann}.} \bibinfo{year}{2018}\natexlab{}.
\newblock \showarticletitle{{Design and construction of curved support
  structures with repetitive parameters}}. In
  \bibinfo{booktitle}{\emph{Advances in Architectural Geometry (AAG) 2018}}.
\newblock
\urldef\tempurl%
\url{https://www.semanticscholar.org/paper/Design-and-construction-of-curved-support-with-Schling-Kilian/1f71181f71ca66ab0f9347a4d7bbda08dc246dc9}
\showURL{%
\tempurl}


\bibitem[\protect\citeauthoryear{Shukhov}{Shukhov}{1896}]%
        {Shukhov1896}
\bibfield{author}{\bibinfo{person}{Vladimir Shukhov}.}
  \bibinfo{year}{1896}\natexlab{}.
\newblock \bibinfo{title}{{Rotunda of the Panrussian Exposition (Nizhny
  Novgorod, 1896) | Structurae}}.
\newblock
\newblock
\urldef\tempurl%
\url{https://structurae.net/en/structures/rotunda-of-the-panrussian-exposition}
\showURL{%
\tempurl}


\bibitem[\protect\citeauthoryear{Soriano}{Soriano}{2017}]%
        {Soriano2017}
\bibfield{author}{\bibinfo{person}{Enrique Soriano}.}
  \bibinfo{year}{2017}\natexlab{}.
\newblock \showarticletitle{{Low-Tech Geodesic Gridshell: Almond Pavilion}}.
\newblock \bibinfo{journal}{\emph{archidoct}}  \bibinfo{volume}{4}
  (\bibinfo{year}{2017}), \bibinfo{pages}{29}.
\newblock


\bibitem[\protect\citeauthoryear{Soriano, Sastre, and Boixader}{Soriano
  et~al\mbox{.}}{2019}]%
        {Soriano2019}
\bibfield{author}{\bibinfo{person}{Enrico Soriano}, \bibinfo{person}{Ramon
  Sastre}, {and} \bibinfo{person}{Dionis Boixader}.}
  \bibinfo{year}{2019}\natexlab{}.
\newblock \showarticletitle{{G-shells: Flat collapsible geodesic mechanisms for
  gridshells}}. In \bibinfo{booktitle}{\emph{IASS Annual Symposium 2019 –
  Structural Membranes}}. \bibinfo{address}{Barcelona}.
\newblock


\bibitem[\protect\citeauthoryear{Stein, Grinspun, and Crane}{Stein
  et~al\mbox{.}}{2018}]%
        {Stein2018}
\bibfield{author}{\bibinfo{person}{Oded Stein}, \bibinfo{person}{Eitan
  Grinspun}, {and} \bibinfo{person}{Keenan Crane}.}
  \bibinfo{year}{2018}\natexlab{}.
\newblock \showarticletitle{{Developability of triangle meshes}}.
\newblock \bibinfo{journal}{\emph{ACM Transactions on Graphics}}
  \bibinfo{volume}{37}, \bibinfo{number}{4} (\bibinfo{date}{Aug.}
  \bibinfo{year}{2018}), \bibinfo{pages}{1--14}.
\newblock
\showISSN{0730-0301}
\urldef\tempurl%
\url{https://doi.org/10.1145/3197517.3201303}
\showDOI{\tempurl}


\bibitem[\protect\citeauthoryear{Surazhsky, Surazhsky, Kirsanov, Gortler, and
  Hoppe}{Surazhsky et~al\mbox{.}}{2005}]%
        {surazhsky05}
\bibfield{author}{\bibinfo{person}{Vitaly Surazhsky}, \bibinfo{person}{Tatiana
  Surazhsky}, \bibinfo{person}{Danil Kirsanov}, \bibinfo{person}{Steven~J.
  Gortler}, {and} \bibinfo{person}{Hugues Hoppe}.}
  \bibinfo{year}{2005}\natexlab{}.
\newblock \showarticletitle{Fast Exact and Approximate Geodesics on Meshes}.
\newblock \bibinfo{journal}{\emph{ACM Transactions on Graphics}}
  \bibinfo{volume}{24}, \bibinfo{number}{3} (\bibinfo{date}{July}
  \bibinfo{year}{2005}), \bibinfo{pages}{553–560}.
\newblock
\showISSN{0730-0301}
\urldef\tempurl%
\url{https://doi.org/10.1145/1073204.1073228}
\showDOI{\tempurl}


\bibitem[\protect\citeauthoryear{Takezawa, Imai, Shida, and Maekawa}{Takezawa
  et~al\mbox{.}}{2016}]%
        {Takezawa2016a}
\bibfield{author}{\bibinfo{person}{Masahito Takezawa}, \bibinfo{person}{Takuma
  Imai}, \bibinfo{person}{Kentaro Shida}, {and} \bibinfo{person}{Takashi
  Maekawa}.} \bibinfo{year}{2016}\natexlab{}.
\newblock \showarticletitle{{Fabrication of freeform objects by principal
  strips}}.
\newblock \bibinfo{journal}{\emph{ACM Transactions on Graphics}}
  \bibinfo{volume}{35}, \bibinfo{number}{6} (\bibinfo{date}{Nov.}
  \bibinfo{year}{2016}), \bibinfo{pages}{1--12}.
\newblock
\showISSN{07300301}
\urldef\tempurl%
\url{https://doi.org/10.1145/2980179.2982406}
\showDOI{\tempurl}


\bibitem[\protect\citeauthoryear{Tang, Sun, Gomes, Wallner, and Pottmann}{Tang
  et~al\mbox{.}}{2014}]%
        {Tang2014a}
\bibfield{author}{\bibinfo{person}{Chengcheng Tang}, \bibinfo{person}{Xiang
  Sun}, \bibinfo{person}{Alexandra Gomes}, \bibinfo{person}{Johannes Wallner},
  {and} \bibinfo{person}{Helmut Pottmann}.} \bibinfo{year}{2014}\natexlab{}.
\newblock \showarticletitle{{Form-finding with polyhedral meshes made simple}}.
\newblock \bibinfo{journal}{\emph{ACM Transactions on Graphics}}
  \bibinfo{volume}{33}, \bibinfo{number}{4} (\bibinfo{date}{July}
  \bibinfo{year}{2014}), \bibinfo{pages}{1--9}.
\newblock
\showISSN{07300301}
\urldef\tempurl%
\url{https://doi.org/10.1145/2601097.2601213}
\showDOI{\tempurl}


\bibitem[\protect\citeauthoryear{Vekhter, Zhuo, Fandino, Huang, and
  Vouga}{Vekhter et~al\mbox{.}}{2019}]%
        {Vekhter2019}
\bibfield{author}{\bibinfo{person}{Josh Vekhter}, \bibinfo{person}{Jiacheng
  Zhuo}, \bibinfo{person}{Luisa F~Gil Fandino}, \bibinfo{person}{Qixing Huang},
  {and} \bibinfo{person}{Etienne Vouga}.} \bibinfo{year}{2019}\natexlab{}.
\newblock \showarticletitle{{Weaving geodesic foliations}}.
\newblock \bibinfo{journal}{\emph{ACM Transactions on Graphics}}
  \bibinfo{volume}{38}, \bibinfo{number}{4} (\bibinfo{date}{July}
  \bibinfo{year}{2019}), \bibinfo{pages}{1--22}.
\newblock
\showISSN{07300301}
\urldef\tempurl%
\url{https://doi.org/10.1145/3306346.3323043}
\showDOI{\tempurl}


\bibitem[\protect\citeauthoryear{Voss}{Voss}{1907}]%
        {Voss}
\bibfield{author}{\bibinfo{person}{Aurel Voss}.}
  \bibinfo{year}{1907}\natexlab{}.
\newblock \showarticletitle{{Über diejenigen Flächen, welche durch zwei
  Scharen von Kurven konstanter geodätischer Krümmung in infinitesimale
  Rhomben zerlegt werden}}.
\newblock \bibinfo{series}{Sitzungsbericht der Bayerischen Akademie der
  Wissenschaften}, Vol.~\bibinfo{volume}{36,7}.
  \bibinfo{publisher}{Verl.d.K.B.Akad.d.Wiss.}, \bibinfo{address}{München}.
\newblock
\urldef\tempurl%
\url{http://publikationen.badw.de/de/003388868}
\showURL{%
\tempurl}


\bibitem[\protect\citeauthoryear{Vouga, H{\"{o}}binger, Wallner, and
  Pottmann}{Vouga et~al\mbox{.}}{2012}]%
        {Vouga2012a}
\bibfield{author}{\bibinfo{person}{Etienne Vouga}, \bibinfo{person}{Mathias
  H{\"{o}}binger}, \bibinfo{person}{Johannes Wallner}, {and}
  \bibinfo{person}{Helmut Pottmann}.} \bibinfo{year}{2012}\natexlab{}.
\newblock \showarticletitle{{Design of self-supporting surfaces}}.
\newblock \bibinfo{journal}{\emph{ACM Transactions on Graphics}}
  \bibinfo{volume}{31}, \bibinfo{number}{4} (\bibinfo{date}{July}
  \bibinfo{year}{2012}), \bibinfo{pages}{1--11}.
\newblock
\showISSN{07300301}
\urldef\tempurl%
\url{https://doi.org/10.1145/2185520.2185583}
\showDOI{\tempurl}


\bibitem[\protect\citeauthoryear{Wallner, Schiftner, Kilian, Fl{\"{o}}ry,
  H{\"{o}}binger, Deng, Huang, and Pottmann}{Wallner et~al\mbox{.}}{2010}]%
        {Wallner2010}
\bibfield{author}{\bibinfo{person}{Johannes Wallner},
  \bibinfo{person}{Alexander Schiftner}, \bibinfo{person}{Martin Kilian},
  \bibinfo{person}{Simon Fl{\"{o}}ry}, \bibinfo{person}{Mathias
  H{\"{o}}binger}, \bibinfo{person}{Bailin Deng}, \bibinfo{person}{Qixing
  Huang}, {and} \bibinfo{person}{Helmut Pottmann}.}
  \bibinfo{year}{2010}\natexlab{}.
\newblock \showarticletitle{{Tiling Freeform Shapes With Straight Panels:
  Algorithmic Methods.}}
\newblock In \bibinfo{booktitle}{\emph{Advances in Architectural Geometry
  2010}}. \bibinfo{publisher}{Springer Vienna}, \bibinfo{address}{Vienna},
  \bibinfo{pages}{73--86}.
\newblock
\urldef\tempurl%
\url{https://doi.org/10.1007/978-3-7091-0309-8_5}
\showDOI{\tempurl}


\bibitem[\protect\citeauthoryear{Wang, Pellis, Rist, Pottmann, and
  M{\"{u}}ller}{Wang et~al\mbox{.}}{2019}]%
        {Wang2019}
\bibfield{author}{\bibinfo{person}{Hui Wang}, \bibinfo{person}{Davide Pellis},
  \bibinfo{person}{Florian Rist}, \bibinfo{person}{Helmut Pottmann}, {and}
  \bibinfo{person}{Christian M{\"{u}}ller}.} \bibinfo{year}{2019}\natexlab{}.
\newblock \showarticletitle{{Discrete geodesic parallel coordinates}}.
\newblock \bibinfo{journal}{\emph{ACM Transactions on Graphics}}
  \bibinfo{volume}{38}, \bibinfo{number}{6} (\bibinfo{date}{Nov.}
  \bibinfo{year}{2019}), \bibinfo{pages}{1--13}.
\newblock
\urldef\tempurl%
\url{https://doi.org/10.1145/3355089.3356541}
\showDOI{\tempurl}


\end{thebibliography}

\end{document}